\documentclass[aps,prx,twocolumn,longbibliography,superscriptaddress]{revtex4-2}

\usepackage{graphicx}
\usepackage{dcolumn}
\usepackage{bm}
\usepackage[english]{babel}
\usepackage{lipsum} 
\usepackage{times}
\usepackage{amsfonts,amssymb,bm}
\usepackage{amsmath}
\usepackage{epsfig}
\usepackage{mathrsfs}
\usepackage{color,soul}
\usepackage{bbold}
\usepackage{ulem}
\usepackage{hyperref}
\hypersetup{colorlinks=true,linktoc=all,linkcolor=blue,breaklinks=true,citecolor=magenta,urlcolor=blue}
\usepackage[english]{babel}
\usepackage{pstricks}
\usepackage{multirow}
\usepackage{textcomp}
\usepackage{braket}
\usepackage{empheq}
\usepackage{xcolor}
\usepackage{rotating}
\usepackage{amsthm}
\usepackage{siunitx}
\usepackage{eurosym}
\usepackage[babel]{csquotes}

\usepackage{mwe}

\newtheorem*{theorem}{Theorem}
\usepackage[framemethod=default]{mdframed}
\usepackage{newfloat}

\DeclareFloatingEnvironment[name={S}]{suppfigure}
\DeclareFloatingEnvironment[name={S}]{supptable}

\newmdenv[linecolor=white,backgroundcolor=gray!15!]{myframe}

\usepackage{dsfont}
\usepackage[tight]{subfigure}
\usepackage{verbatim}
\usepackage{units}
\usepackage{natbib}
\usepackage{enumitem}
\usepackage{mathrsfs}
\usepackage{leftidx}
\usepackage{xspace}

\definecolor{bazaar}{rgb}{0.6, 0.47, 0.48}
\definecolor{green1}{rgb}{0.33, 0.7, 0.69}
\definecolor{grey1}{rgb}{0.51, 0.54, 0.51}

\colorlet{RED}{red}

\begin{document}

\title{Inferring work by quantum superposing forward and time-reversal evolutions}

\author{Giulia Rubino}
\thanks{giulia.rubino@univie.ac.at}
\affiliation{University of Vienna, Faculty of Physics,
Vienna Center for Quantum Science and Technology (VCQ), Boltzmanngasse 5, Vienna 1090, Austria}
\affiliation{Quantum Engineering Technology Labs, H. H. Wills Physics
Laboratory and Department of Electrical \& Electronic Engineering,
University of Bristol, Bristol BS8 1FD, United Kingdom}

\author{Gonzalo Manzano}
\affiliation{Institute for Quantum Optics \& Quantum Information (IQOQI), Austrian Academy of Sciences\symbol{44} Boltzmanngasse 3\symbol{44} Vienna 1090\symbol{44} Austria}
\affiliation{Institute for Cross-Disciplinary Physics and Complex Systems (IFISC) UIB-CSIC, Campus Universitat Illes Balears, E-07122 Palma de Mallorca, Spain}

\author{Lee A. Rozema}
\affiliation{University of Vienna, Faculty of Physics,
Vienna Center for Quantum Science and Technology (VCQ), Boltzmanngasse 5, Vienna 1090, Austria}

\author{Philip Walther}
\affiliation{University of Vienna, Faculty of Physics,
Vienna Center for Quantum Science and Technology (VCQ), Boltzmanngasse 5, Vienna 1090, Austria}
\affiliation{University of Vienna, Research Platform TURIS, Boltzmanngasse 5, Vienna 1090, Austria}

\author{Juan M. R. Parrondo}
\affiliation{Departamento de Física Atómica\symbol{44} Molecular y Nuclear and GISC, Universidad Complutense Madrid, 28040 Madrid, Spain}

\author{\v{C}aslav Brukner}
\thanks{caslav.brukner@univie.ac.at}
\affiliation{University of Vienna, Faculty of Physics,
Vienna Center for Quantum Science and Technology (VCQ), Boltzmanngasse 5, Vienna 1090, Austria}
\affiliation{Institute for Quantum Optics \& Quantum Information (IQOQI), Austrian Academy of Sciences\symbol{44} Boltzmanngasse 3\symbol{44} Vienna 1090\symbol{44} Austria}


\date{\today}
\begin{abstract}
The study of thermodynamic fluctuations allows one to relate the free energy difference between two equilibrium states with the work done on a system through processes far from equilibrium. This finding plays a crucial role in the quantum regime, where the definition of work becomes non-trivial. Based on these relations, here we develop 
a simple interferometric 
method  allowing a direct estimation of the work distribution and the average dissipative work during a driven thermodynamic process by superposing the forward and time-reversal evolutions of the process. We show that our scheme provides useful upper bounds on the average dissipative work  even 
without full control over the thermodynamic process, and we propose methodological variations depending on the possible experimental limitations encountered. Finally, we exemplify its applicability by an experimental proposal for implementing our method on a quantum photonics system, on which the thermodynamic process 
is performed 
through polarization rotations induced by liquid crystals acting in a discrete temporal regime.
\end{abstract}

\maketitle


\section{Introduction}

While microscopic dynamical physical laws of both classical and quantum physics are time-symmetric, and hence reversible, the dynamics of macroscopic quantities exhibit a preferred temporal direction.
The physical law formalizing this concept is the second law of thermodynamics, whereby the ``arrow of time'' \cite{Eddington_1928} is associated with a production of  entropy~\cite{Callen1985}. According to this law, for instance, if we take a vessel divided by a wall, and put a gas in only one half of the vessel, when we remove the wall we will observe with a near-unity probability the gas expanding and occupying the whole vessel. Because of its unidirectional temporal evolution, this phenomenon has often been used to differentiate between past and future. There is, however, a non-zero probability that at a time all the molecules may happen to visit one half of the vessel. 
In this regard, the development of so-called ``fluctuation theorems'', both for classical \cite{EvansPRL, JarzynskiPRL, CrooksPRE, JarzynskiREV, SeifertREV} and quantum \cite{RevModPhys.81.1665, RevModPhys.83.771, Funo2018, Chetrite:2012, PhysRevE.88.032146, PhysRevE.89.012127, PhysRevE.92.032129, PhysRevX.6.041017, Iyoda:2017, PhysRevX.8.011019, PhysRevX.8.031037} systems, has led to the sharpening of our understanding of the second law as a statistical law, where the entropy of a system away from equilibrium can spontaneously decrease rather than increase with non-zero probability. As specified by those theorems, the ratio between the probability of entropy-decreasing events and that of entropy-increasing ones vanishes exponentially with the size of the fluctuations, and can hence be neglected in the macroscopic limit~\cite{JarzynskiREV}.

The fundamental and empirical basis for the study of entropy production and thermodynamic irreversibility in driven systems is typically provided by the notion of dissipative work, $W_\mathrm{diss} \equiv W - \Delta F$ (namely, the work invested in a thermodynamic transformation between equilibrium states having a free energy difference $\Delta F$, which cannot be recovered by reversing the driving protocol) ~\cite{JarzynskiPRL, CrooksPRE, Kawai2007, Parrondo2009, LutzPRL}. The fluctuations of the dissipative work in the process can be characterized by constructing the work probability distribution, $P(W)$, associated to the observation of a particular value of $W$ in a single realization of the driving protocol. Such fluctuations are constrained by a refined version of the second law: namely, \textit{Crook's fluctuation theorem}, according to which
\begin{equation} \label{eqn:crooks}
    \frac{P(W)}{\tilde{P}(-W)} = e^{\beta W_\mathrm{diss}},
\end{equation}
where $\tilde{P}(-W)$ is the probability of performing a work $W$ in the time-reversal dynamics, $\beta = 1/k_B T$ is the inverse temperature of the surrounding thermal environment, and $k_B$ is the Boltzmann constant. According to Eq.~\eqref{eqn:crooks}, this probability ratio decreases exponentially with the amount of dissipative work, $W_\mathrm{diss}$, in the realization. Furthermore, Eq.~\eqref{eqn:crooks} implies the famous Jarzynski equality $\langle e^{- \beta W_\mathrm{diss}} \rangle = 1$, where the brackets denote the statistical average with respect to $P(W)$. Jarzynski's equality has severe implications by itself, such as the exponential decay of the probability to observe negative values of $W_\mathrm{diss}$ in the forward dynamics (explicitly, $P(W_\mathrm{diss}< -\zeta) \leq e^{-\beta \zeta}$ for any $\zeta \geq 0$) ~\cite{JarzynskiREV}. 

Work fluctuations have been measured in small classical systems leading to both testing the Crook's theorem and the Jarzynski equality, and developing applications like measurements of free-energy~\cite{Wang2002, Liphardt2002, Ritort2005, PhysRevLett.97.050602, Ueda2010, NatNano.9.358}. In quantum physics, since work is not associated to any observable~\cite{PhysRevE.75.050102}, its definition becomes more complex, and it usually demands the use of the so-called ``two-point measurement (TPM) scheme''~\cite{ RevModPhys.83.771}. 
In the TPM scheme, work is represented as the difference between the initial and final energies of the system, obtained by performing two projective measurements of the Hamiltonian at the beginning and at the end of the forward as well as of the time-reversal process. Extensions to non-ideal measurements~\cite{Watanabe_2014,Talkner_2019,Debarba_2019} and variants of the TPM scheme~\cite{Solinas:2016,PhysRevLett.118.070601,PhysRevLett.120.040602,Sone_2020,Beyer_2020,Micadei_2020} have been also considered recently. The TPM
 approach has been directly implemented in several experiments~\cite{Kim2015, PhysRevLett.120.010601, PhysRevLett.121.088901, Zhang_2018, Wueaav4944}.
However, since implementing projective energy measurements before and after an arbitrary process may be challenging in certain experimental scenarios, and the measurement might annihilate 
the system measured, alternative methods for extracting the work distribution were proposed to circumvent this requirement. For example, in Refs.~\cite{PhysRevLett.110.230601, PhysRevLett.110.230602}, a scheme based on Ramsey interferometry using a single probe qubit was proposed, and subsequently implemented~\cite{Serra2014, Serra2015}, to extract the characteristic function of work in an NMR platform. A similar method to sample the work probability distribution from a generalized measurement scheme was introduced in Refs.~\cite{Paz2014,Chiara_2015,Talkner_2016}, and tested experimentally on an ensemble of cold atoms~~\cite{Cersiola2017}. 
Despite their many advantages and proven efficacy, previous
 schemes often involve indirect measurements requiring post-processing of data, or experimentally demanding entangling operations. 
Developing new accurate and simple methods to directly estimate the work probability distribution and irreversibility (thus refraining from the TPM scheme) is therefore of prime interest in quantum thermodynamics.

In this paper, we 
propose a simple interferometric method for quantifying the work distribution and the average dissipative work associated to a given driving protocol $\Lambda$ during a thermodynamic process. The method enables one to directly read out the relevant transition probabilities between eigenstates of the initial and the final Hamiltonians, which are needed to build the work probability distribution and the relative entropy (or Kullback-Leibler divergence) between the density operators in forward and time-reversal dynamics.
Remarkably, our method requires no entangling operations with separate auxiliary systems, no measurement of the thermodynamic system, no data post-processing, and it runs twice as fast as running the complete protocol $\Lambda$. 
More precisely, in the proposed method we superpose two interferometric paths: along one path the system is driven following the first half of the driving protocol $\Lambda$ (i.e., from $t=0$ to $t=\tau/2$), while along the other path the system is affected by the time-reversal version of the second half of the protocol (from $t=\tau/2$ to  $t=\tau$).
We show that the fringe visibility in the interferometer allows one to quantify both 
the full work probability distribution associated to an arbitrary protocol $\Lambda$, and the relative entropy between the states in the forward and the time-reversal dynamics at any instant of time, assuming that the initial and final Hamiltonians are known. Moreover, in the case of limited control over preparations, our scheme still provides useful upper bounds on the average dissipative work. 

Since single photons provide advantages in interferometric schemes due to their robustness, individual addressability and the intrinsic mobility, we  propose a photonic implementation of 
our scheme 
where the Hamiltonian of the thermodynamic system is represented by 
the polarization of a single photon. Other platforms 
that may be used to realize the scheme include ultracold atoms~\cite{Chiara_2015,Cersiola2017}, or NMR spectroscopy of nuclear spins~\cite{Serra2014, Serra2015}. Also a methodologically related scheme was proposed recently to investigate the thermodynamic arrow of time in a quantum superposition of the forward and time-reversal processes~\cite{Rubino:2020}.



\section{PROCEDURE OVERVIEW}



Consider a thermodynamic system $S$ that is driven by a time-dependent Hamiltonian $H(\Lambda(t))$ depending on some control parameter $\Lambda(t)$ which varies from $t=0$ to $t=\tau$, according to a protocol $\Lambda = \{\Lambda(t) : 0 \leq t \leq \tau \}$. The system starts the evolution in a thermal state ${\rho}_0^{\text{th}}= \exp[{-\beta (H_0 - F_0)}]$ in equilibrium with a thermal reservoir at inverse temperature $\beta$, where $F_0$ is the free energy corresponding to the initial Hamiltonian $H_0 \equiv H\bigl(\Lambda(0)\bigr)$. The system is then isolated from the environment, and the driving protocol $\Lambda$ is applied, bringing the system to an out-of-equilibrium state $\rho (t) = U(t,0) \, \rho_0^{\text{th}} \, U^\dagger(t,0)$, where $U(t,0) = \mathbf{\overrightarrow{T}}\exp[{-\frac{i}{\hbar}\int_0^{t} H\bigl(\Lambda(t')\bigr) dt'}]$, $\mathbf{\overrightarrow{T}}$ being the so-called ``time-ordering'' operator resulting from the Dyson decomposition. Once the driving protocol is ended at time $\tau$, the system may eventually equilibrate again from $\rho_\tau = \rho(\tau)$  to the reservoir temperature, thereby reaching the thermal state $\rho_{\tau}^{\text{th}}=\exp\bigl[{-\beta (H_\tau - F_\tau)}\bigr]$, corresponding to the final Hamiltonian $H_\tau \equiv H(\Lambda(\tau))$ and the free energy $F_\tau$.

Together with the above thermodynamic process, we consider its time-reversal twin. In the reverse process, the system starts the evolution at time $t=0$ with Hamiltonian $\Theta H_\tau \Theta^\dagger$ in equilibrium with the thermal reservoir, that is,  $\tilde{\rho}_0^{\mathrm{th}} \equiv \Theta \rho_{\tau}^{\text{th}} \Theta^\dagger =  \exp\bigl[{-\beta ( \Theta H_\tau \Theta^\dagger - F_\tau)}\bigr]$. Here, $\Theta$ is the (anti-unitary) time-reversal operator, responsible for changing the sign of observables with odd parity (such as momentum, or spin under time-reversal). The time-reversal operator fulfills $\Theta \, \, \mathbb{1}i = - \mathbb{1}i \, \Theta$ and $\Theta \, \Theta^\dagger = \Theta^\dagger \, \Theta = \mathbb{1}$. 
The system is then driven according to the time-reversal protocol $\tilde{\Lambda} = \{ \tilde{\Lambda}(t) \equiv \Lambda(\tau -t): 0\leq t \leq \tau \}$, corresponding to the inverse sequence of values of the control parameter. This brings the system out-of-equilibrium to the state $\tilde{\rho}(t) = \tilde{U}(t,0) \, \tilde{\rho}_0^{\text{th}} \, \tilde{U}^\dagger(t,0)$ at intermediate times, where now $\tilde{U}(t,0) = \mathbf{\overrightarrow{T}}\exp\bigl[{-\frac{i}{\hbar}\int_0^{t} \Theta H\bigl(\tilde{\Lambda}(t') \bigr) \Theta^\dagger dt'}\bigr]$. After completing the protocol $\tilde{\Lambda}$, the system may return back to equilibrium at time $t = \tau$, reaching $\tilde{\rho}_{\tau}^{\text{th}}= \Theta \, \rho_0^{\text{th}} \, \Theta^\dagger = \exp[{-\beta (\Theta H_0 \Theta^\dagger - F_0)}]$.  

\begin{figure}
\centering
\includegraphics[width=0.9\columnwidth]{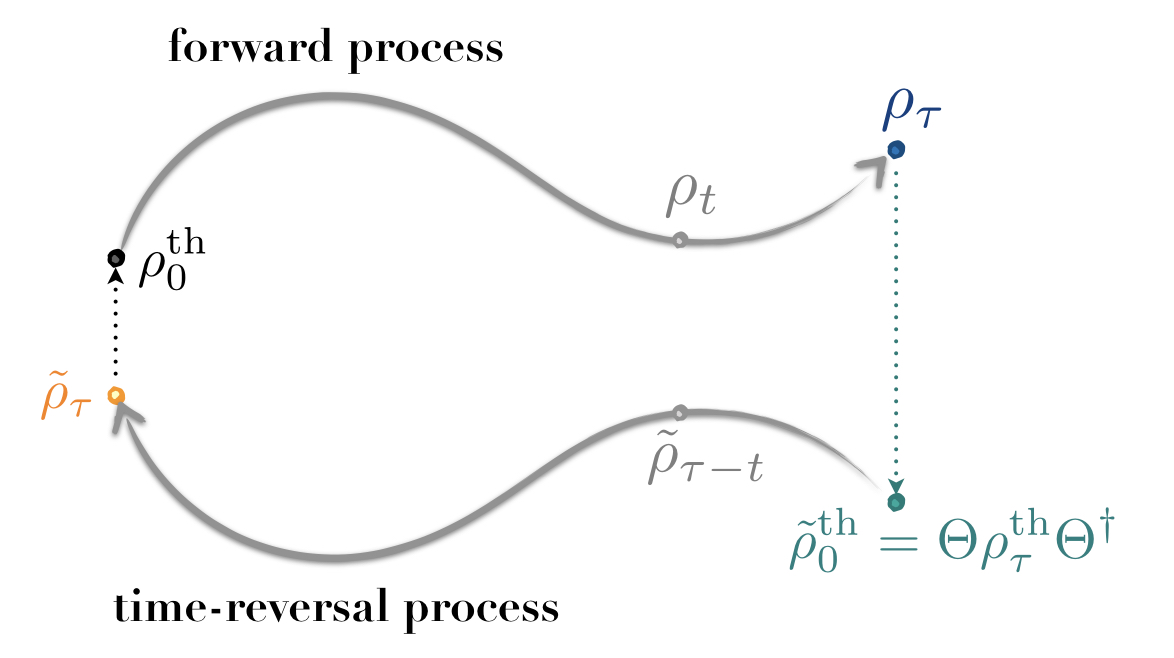}
\caption{Diagrammatic representation of the forward and time-reversal evolutions of the thermodynamic system. An initial thermal state $\rho_0^{\text{th}}$ with Hamiltonian $H_0$ is driven into a final, non-equilibrium state $\rho_{\tau}$. It then eventually equilibrates at the reservoir temperature, reaching the thermal state $\rho_{\tau}^{\text{th}}$. (If the driving process was reversible, quasi-static, the system would have ended in the state $\rho_{\tau}^{\text{th}},$ immediately after the drive.) Along the driving process the Hamiltonian is changed from $H_0$ to $H_{\tau}$. Analogously, in the process' time-reversal twin a thermal state $\tilde{\rho}_0^{\text{th}}=\rho_{\tau}^{\text{th}}$ with Hamiltonian $H_{\tau}$ evolves into a state $\tilde{\rho}_{\tau}$ and then it eventually equilibrates to the state $\tilde{\rho}_{\tau}^{\text{th}}=\rho_0^{\text{th}}$.}
\label{img:ThermoIrreversibility1}
\end{figure}

We denote by $\bigl\vert{E_n^{(0)}}\bigr\rangle$ the initial energy eigenstates of the system in the forward process, and by $p_n^{(0)} = e^{-\beta(E_n^{(0)} - F_0)}$ the probability that the system has energy $E_n^{(0)}$. Analogously, the initial eigenstates of the system in the time-reversal process read $ \Theta \bigl\vert{E_m^{(\tau)}}\bigr\rangle$, with $\tilde{p}_m^{(0)} = e^{-\beta(E_m^{(\tau)} - F_\tau)}$ being the corresponding probabilities to measure the energy $E_m^{(\tau)}$. The work probability distribution in the TPM scheme results then~\cite{RevModPhys.83.771}:
\begin{equation} \label{eqn:wprob}
P(W) = \sum_{m, n}  p_n^{(0)}~ p_{m|n} \, \delta\bigl(W - (E_m^{(\tau)} - E_n^{(0)})\bigr),   
\end{equation}
where we introduced the conditional probabilities $p_{m|n} = \bigl\vert\langle E_m^{(\tau)}| \, U(\tau,0)|E_n^{(0)} \rangle\bigl\vert^2$ to find the system in the eigenstate $ \ket{E_m^{(\tau)}}$ in the second projective energy measurement after the unitary evolution $U(\tau,0)$, given that it was found to be in $\ket{E_n^{(0)}}$ in the first measurement. Likewise, the work distribution in the time-reversal process reads $\tilde{P}(W) = \sum_{m, n}  \tilde{p}_m^{(0)}~ \tilde{p}_{n|m} \, \delta\bigl(W - (E_n^{(0)} - E_m^{(\tau)})\bigr)$, for which $\tilde{p}_{n|m} = \Bigl\vert\langle E_n^{(0)}| \Theta^\dagger \tilde{U}(\tau,0)\, \Theta|E_m^{(\tau)}\rangle\Bigr\vert^2$. The {micro-reversibility relation} for non-autonomous systems~\cite{RevModPhys.83.771} reads:
\begin{equation} \label{eqn:microrev}
\Theta^\dagger \, \tilde{U}(\tau-t,0) \, \Theta = U^\dagger(\tau,t).
\end{equation}
Using the micro-reversibility relation \eqref{eqn:microrev}, we obtain $\tilde{p}_{n|m} = p_{m|n}$. This relation is the key property to obtain Crook's theorem in Eq.~\eqref{eqn:crooks}~\footnote{The same result can be extended to Hamiltonians which are not invariant under time-reversal. In such a case, the initial state of the time-reversal process needs to incorporate the broken symmetry, that is, $\tilde{\rho}_0^\mathrm{th}= \Theta \, \rho_\tau^{\mathrm{th}}$}. Furthermore, we assume that the Hamiltonian is invariant under time-reversal [i.e., $\Theta H(t) =  H(t) \Theta$]. As a consequence, the relations $\Theta \ket{E_n^{(0)}} = \ket{E_n^{(0)}}$ and $\Theta \ket{E_m^{(\tau)}} = \ket{E_m^{(\tau)}}$ are also verified.

In Refs.~\cite{Kawai2007, Parrondo2009}, the authors derived an important relation closely connected to Crook's theorem linking the dissipative work produced during the protocol $\Lambda$ with the relative entropy between the density operators in forward and time-reversal dynamics at any intermediate instant of time $t$:
\begin{equation} \label{eqn:basic}
    \beta \langle W_\mathrm{diss} \rangle = S\bigl(\rho(t) \, || \, \Theta^\dagger \tilde{\rho}(\tau -t) \, \Theta\bigr),
\end{equation}
where $S(\rho||\sigma):=\text{Tr}\bigl[\rho \, \text{ln}(\rho) - \rho \, \text{ln}(\sigma) \bigr] \geq 0$ is the relative entropy between two generic states $\rho$ and $\sigma$. Reversible processes for which the state in the forward dynamics is statistically indistinguishable from the one generated in the time-reversal dynamics do not dissipate work, $\langle W_\mathrm{diss} \rangle = 0$, and therefore all the work performed during the protocol $\Lambda$, $\langle W \rangle = \Delta F$, can be recovered back implementing the time-reversal protocol $\tilde{\Lambda}$. Importantly, the equality in Eq.~\eqref{eqn:basic} is obtained in the case of a closed system following unitary dynamics, as in the TPM scheme presented above. For open systems, the equality above is instead replaced by an inequality after tracing out environmental degrees of freedom~\cite{Kawai2007, Parrondo2009}.

In the following, we present an interferometric scheme that allows us to directly measure the conditional probabilities $p_{m|n}$ (and therefore $\tilde{p}_{n|m}$) without implementing the TPM scheme, but resorting to the visibility of fringes in the interferometer. This enables us to construct $P(W)$ and $\tilde{P}(W)$, and hence the relative entropy $S\bigl(\rho(t) \, || \, \Theta^\dagger \tilde{\rho}(\tau -t) \, \Theta\bigr)$ in Eq.~\eqref{eqn:basic}.

\section{Interferometric Scheme}

\begin{figure}
\centering
\includegraphics[width=\columnwidth]{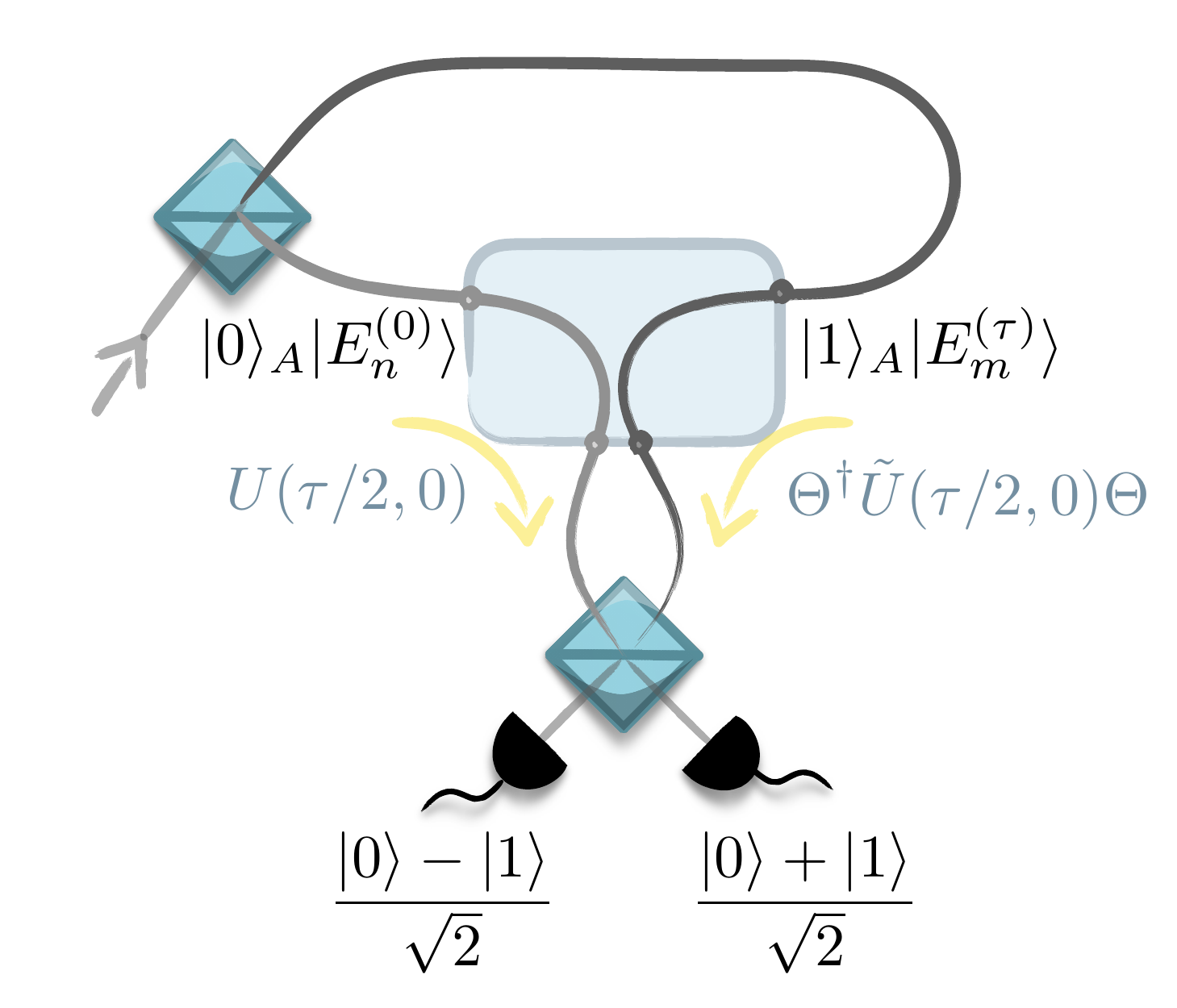}
\caption{Schematic representation of the interferometric technique to directly estimate the work dissipation. A thermodynamic quantum system $S$ is prepared in the state $\vert E_n^{(0)} \rangle$ ($\vert E_m^{(\tau)} \rangle$) when the auxiliary system is in $\ket{0}_{A}$ ($\ket{1}_{A}$) at time $t=0$. 
The operation $U(\tau/2,0)$ is then applied to $S$ when traveling along the path $\ket{0}_{A}$, while the operation $\tilde{U}(\tau/2,0)$ followed by the time-inversion operation $\Theta^\dagger$ is applied to $S$ along the path $\ket{1}_{A}$. The resulting state is as in Eq.~\eqref{eqn:rho-tau2}. The two quantum superposed amplitudes are then interfered with each other, and the auxiliary system is measured in the $\bigl\lbrace \bigl(\ket{0}_A\pm\ket{1}_A\bigr)/\sqrt{2}\bigr\rbrace$ basis. The same setup can be used in the case of a limited preparation, in which case the input state is the thermal state $\rho_0^{\text{th}}$ ($\tilde{\rho}_0^{\text{th}}$) for $\ket{0}_{A}$ ($\ket{1}_{A}$).}
\label{img:ThermoIrreversibility2}
\end{figure}

The main idea of our scheme is to entangle the system of interest with a two-level ``auxiliary system'', and implement different dynamics (forward and time-reversal) on each of the two states of the auxiliary system. To fix ideas, we assume that the auxiliary system is the path of a single photon in a Mach-Zehnder interferometer such as the one depicted in  Fig.~\ref{img:ThermoIrreversibility2}, and denote by $\lbrace \ket{0}_{A}, \ket{1}_{A}\rbrace$  the basis  of the two possible paths.
(We stress, however, that this auxiliary system does not have to be encoded in the path, but can be any degree of freedom which can be suitably controlled.)
Suppose now that, in one of the two states of the superposition (say, $\ket{0}_{A}$), the system is prepared in the state $\ket{E_n^{(0)}}$, while in the other state of the superposition ($\ket{1}_{A}$) the preparation is $\ket{E_m^{(\tau)}}$ for a certain choice of $n$ and $m$. Consequently, the initial state is the pure state
\begin{equation}
  \ket{\psi(0)}_{S,A}=  \frac{1}{\sqrt{2}}
  \left[ \ket{0}_A\ket{E_n^{(0)}}+
  \ket{1}_A\ket{E_m^{(\tau)}}\right]
  \label{eqn:psi0}
\end{equation}
The operation $U(\tau/2, 0)$ is then applied to $S$ in the path $\ket{0}_{A}$, while, on the path $\ket{1}_{A}$, the operation $\tilde{U}(\tau/2, 0)$ is performed, followed by the time-inversion operation $\Theta^\dagger$. The total evolution is given by
\begin{equation}
\ket{0}_{A}\bra{0}_{A} \otimes U(\tau/2,0)+
\ket{1}_{A}\bra{1}_{A} \otimes \Theta^\dagger \tilde U(\tau/2,0).
\end{equation}

Therefore, at the time $\tau/2$, the state of system and path degree of freedom will read:
\begin{align} \label{eqn:rho-tau2}
& \rho_{S,A}(\tau/2)  =  \\
~~ &\dfrac{1}{2} \Bigl\{\ket{0}_{A}\bra{0}_{A} \otimes U(\tau/2,0)  \ket{E_n^{(0)}} \bra{E_n^{(0)}} U^\dagger(\tau/2,0) \notag \\
~~ & + \ket{0}_{A}\bra{1}_{A} \otimes U(\tau/2,0)  \ket{E_n^{(0)}} \bra{E_m^{(\tau)}} \tilde{U}^\dagger(\tau/2,0) \, \Theta  \notag\\
~~ & + \ket{1}_{A}\bra{0}_{A} \otimes \Theta^\dagger \tilde{U}(\tau/2,0) \ket{E_m^{(\tau)}}\bra{E_n^{(0)}} U^\dagger(\tau/2,0)  \notag\\
~~ & + \ket{1}_{A}\bra{1}_{A} \otimes \Theta^\dagger \tilde{U}(\tau/2,0) \ket{E_m^{(\tau)}} \bra{E_m^{(\tau)}} \tilde{U}^\dagger(\tau/2,0) \, \Theta \Bigr\}. \notag
\end{align}
If now we marginalize on the path degree of freedom (i.e., we trace out the thermodynamic system), we obtain
\begin{align} \label{eqn:rho-ancilla}
&\rho_{A}(\tau/2)= \text{Tr}_{S}[\rho_{S,A}(\tau/2)] = \dfrac{1}{2}\Bigl\{\ket{0}_{A} \bra{0}_{A} +~ \ket{1}_{A} \bra{1}_{A} \notag \\
&+ \ket{0}_{A} \bra{1}_{A} \text{Tr}_{S} \Bigl[U(\tau/2,0)  \ket{E_n^{(0)}} \bra{E_m^{(\tau)}} \tilde{U}^\dagger(\tau/2,0) \, \Theta \Bigr] +  \notag\\
&+ \ket{1}_{A} \bra{0}_{A} \text{Tr}_{S} \Bigl[\Theta^\dagger \, \tilde{U}(\tau/2,0) \ket{E_m^{(\tau)}}\bra{E_n^{(0)}} U^\dagger(\tau/2,0) \Bigr]\Bigr\}.
\end{align}

Similarly, if we trace out the auxiliary system, we achieve a mixture between the state of the driven system at time $\tau/2$ in forward and time-reversal processes:
\begin{align}
\rho_S(\tau/2) = \dfrac{1}{2} \Bigl[&\overbrace{ U(\tau/2,0)  \ket{E_n^{(0)}} \bra{E_n^{(0)}} U^\dagger(\tau/2,0)}^{= \rho_n(\tau/2)} +\\
+ &\underbrace{\Theta^\dagger \, \tilde{U}(\tau/2,0) \ket{E_m^{(\tau)}} \bra{E_m^{(\tau)}} \tilde{U}^\dagger(\tau/2,0) \, \Theta}_{= \Theta^\dagger \, \tilde{\rho}_m(\tau/2) \, \Theta} \Bigr].\notag
\end{align}
We see that the state of the system is a mixture of $\rho_n(\tau/2)$ (i.e., the state resulting from the forward evolution during a time interval $\tau/2$ with initial condition $\ket{E_n^{(0)}}$), and
$\tilde\rho_m(\tau/2)$ (which is the state resulting from the time-reversal evolution during a time interval $\tau/2$ with initial condition $\ket{E_m^{(0)}}$).

Ultimately, our aim is to relate the information gained by measuring the output ports of the interferometer to the work statistics and the ``degree of reversibility'' of the thermodynamic processes. 
This degree of reversibility is related to the distinguishability of the two possible paths followed by the auxiliary system in the interferometer. If we measure the final state $\rho_A(\tau/2)$ in the basis $(\ket{0}_A\pm\ket{1}_A)/\sqrt{2}$ (see Fig.~\ref{img:ThermoIrreversibility2}), the probability to get each of the two possible results is
\begin{equation}
p_\pm=\frac{1}{2}\pm \operatorname{Re}\Bigl(\bra{0}_A\rho_A(\tau/2)\ket{1}_A\Bigr).
\end{equation}
In an interferometer, the difference $|p_+-p_-|/2$ is called  interferometric visibility or fringe, and is related to our capacity to identify the path followed by the auxiliary system
\cite{PhysRevLett.77.2154}:
\begin{align}
\mathcal{V}_{m, n}&= \bigl|\bra{0}_A\rho_A(\tau/2)\ket{1}_A\bigr| \nonumber\\
&=\Bigl|\text{Tr}_{S} \Bigl[U(\tau/2,0)  \ket{E_n^{(0)}} \bra{E_m^{(\tau)}} \tilde{U}^\dagger(\tau/2,0) \, \Theta \Bigr] \Bigr|.
\label{eqn:visibility}
\end{align}
Now, we crucially apply the micro-reversibility relation in Eq.~\eqref{eqn:microrev}, to realize that $\Theta^\dagger \, \tilde{U}^\dagger(\tau/2,0) \, \Theta= U(\tau, \tau/2)$. Inserting this into Eq.~\eqref{eqn:visibility}, and using the cyclic property of the trace, we obtain the main result of our proposal:
\begin{align}\label{eqn:result}
\mathcal{V}_{m, n} &=\Bigl|\text{Tr}_{S} \Bigl[U(\tau , 0) \ket{E_n^{(0)}} \bra{E_m^{(\tau)}} \Bigr] \Bigr| \notag \\
&= \Bigl| \bra{E_m^{(\tau)}} U(\tau , 0) \ket{E_n^{(0)}} \Bigr| = \sqrt{p_{m|n}},
\end{align}
where, in the last equality, we identified the expression of the conditional probabilities $p_{m|n}$ of the TPM scheme, and where we identified $U(\tau, \tau/2) \, U(\tau/2,0) = U(\tau, 0)$.

Running this scheme for the $N^2$ different initial states, $n, m = 1, 2, ..., N$ (where $N$ is the dimension of the system Hilbert space), and assuming that we know the eigenenergies $E_n^{(0)}$, $E_m^{(\tau)}$, and the equilibrium free energies $F_0$ and $F_\tau$ (or, equivalently, the initial probabilities $p_n^{(0)}$ and $\tilde{p}_m^{(0)}$), we can readily reconstruct the full probability distribution in Eq.~\eqref{eqn:wprob}:
\begin{equation} \label{eqn:workprob}
P(W) = \sum_{m, n} p_n^{(0)} \mathcal{V}_{m, n}^2 \delta\bigl(W - (E_m^{(\tau)} - E_n^{(0)})\bigr),  
\end{equation}
and its time-reversal twin $\tilde{P}(W)$. We notice that in practice only $(N-1)^2$ of the $N^2$ initial preparations need to be considered, since the properties of the conditional probability imply $\sum_m  \mathcal{V}_{m, n}^2 =1$ for all $n = 1, ..., N$, and for any unital process $p_{m|n}$ becomes doubly stochastic thus we also have $\sum_n  \mathcal{V}_{m, n}^2 =1$ for all $m = 1, ..., N$, as also noticed in Ref.~\cite{Cersiola2017}. 
Furthermore, we can rewrite the r.h.s. of Eq.~\eqref{eqn:basic} 
in terms of known quantities:
\begin{align} \label{eqn:avwork}
& \beta \langle W_\mathrm{diss} \rangle = S\bigl(\rho(t) \, || \, \Theta^\dagger \tilde{\rho}(\tau -t) \, \Theta\bigr) \notag \\ 
&= \sum_n p_n^{(0)} \log p_n^{(0)} - \sum_{m, n} p_n^{(0)} \mathcal{V}_{m, n}^2 \log \tilde{p}_m^{(0)},
\end{align}
which can be alternatively obtained from the average of the work probability distribution in Eq.~\eqref{eqn:workprob}, $\langle W \rangle = \int_{-\infty}^\infty W P(W) \, dW$, and the free energy difference between the initial equilibrium states, $\Delta F = F_\tau - F_0$. As a consequence, this scheme allows, through Eqs.~\eqref{eqn:workprob} and \eqref{eqn:avwork}, the direct estimation of the work dissipation, and the testing of the Jarzynski equality.

\section{Limited preparation and bounds on work dissipation}
\label{sec:lim_prep}

In the previous section, we assumed that we have the ability to prepare a superposition of pairs of energy eigenstates of the initial and final Hamiltonians of the system. Nonetheless, it could be the case that, due to technical limitations, one may not be able to prepare these pure states in the laboratory. For instance, if we do not have full control over the system in its preparation stage, and cannot isolate it from the reservoir, we may only be able to prepare the thermal states $\rho_0^\mathrm{th}$ and $\rho_\tau^\mathrm{th}$. In the following we explore what we can still learn about the work dissipation by exploiting our interferometric scheme in such a situation. We anticipate that, although the full work probability distribution is no longer recoverable in this case, we are still able to provide useful upper bounds on the dissipative work done in the process.

As before, we prepare our auxiliary degree of freedom in a quantum superposition $\frac{1}{\sqrt{2}}\bigl(\ket{0}_{A} + \ket{1}_{A} \bigr)$ at $t<0$. The initial states for the system in the two branches will now be, in general, the mixed thermal states  $\rho_0^{\mathrm{th}}$ and $\rho_\tau^{\mathrm{th}}$. However, hereafter we will make use of their ``purifications'', which can be considered as useful mathematical tools, and may correspond physically to all the environmental degrees of freedom $E$, such that the overall joint state of the system and these degrees of freedom is pure. (Notice that here the environment includes, but  is not limited to, the thermal reservoir. Furthermore, our scheme does not require to have access to the environmental degrees of freedom.) We denote the purifications of the thermal states, respectively, as $\ket{\psi^{(0)}}_{S,E}$ and  $\ket{\tilde{\psi}^{(0)}}_{S,E}$, and they verify $\text{Tr}_{E}\bigl[\ket{\psi^{(0)}}_{S,E} \bra{\psi^{(0)}}_{S,E}\bigr] = \rho_0^{\text{th}}$ and $\text{Tr}_{E}\bigl[\ket{\tilde{\psi}^{(0)}}_{S,E} \bra{\tilde{\psi}^{(0)}}_{S,E}\bigr] = \tilde{\rho}_0^{\text{th}} = \rho_\tau^{\text{th}}$. 

Again, we perform the operation $U(\tau/2, 0)$ in the path $\ket{0}_{A}$ according to the protocol $\Lambda$, and $\tilde{U}(\tau/2, 0)$ in the path $\ket{1}_{A}$ according to $\tilde{\Lambda}$, followed by $\Theta^\dagger$. Notice that the unitaries $U(\tau/2, 0)$ and $\tilde{U}(\tau/2, 0)$ only act on the system of interest, with no effect on the environment. We can then compute the global state of the system, the environment and the auxiliary system at $\tau/2$ similarly as before, and obtain the marginal states for the auxiliary degree of freedom and the composite system consisting of the system and environment. For the latter, we obtain a mixture over the states of the system and the environment at $\tau/2$ in the forward and time-reversal dynamics:
\begin{align}
\label{eqn:detector_states}
\rho_{S,E}(\tau/2) = \dfrac{1}{2} \Bigl[ \rho_{S,E}^{(+)} + \rho_{S,E}^{(-)} \Bigr],
\end{align}
where
\begin{subequations}
\begin{align}
    &\rho_{S,E}^{(+)} =\Bigl(U(\tau/2,0) \otimes \mathbb{1}_E\Bigr)\ket{\psi^{(0)}}_{S,E} \bra{\psi^{(0)}}_{S,E} \notag \\
    &\qquad \qquad \Bigl(U^\dagger(\tau/2,0) \, \otimes \, \mathbb{1}_E\Bigr),\\
    &\rho_{S,E}^{(-)} = \Bigl(\Theta^\dagger \, \tilde{U}(\tau/2,0) \otimes \mathbb{1}_E\Bigr)\ket{\tilde{\psi}^{(0)}}_{S,E} \bra{\tilde{\psi}^{(0)}}_{S,E} \notag \\
    &\qquad \qquad \Bigl(\tilde{U}^\dagger(\tau/2,0) \, \Theta \otimes \, \mathbb{1}_E\Bigr).
\end{align}
\end{subequations}
The corresponding state of the system only will be then an equal probability mixture of the states $\rho_S(\tau/2)= \text{Tr}_E\bigl[ \rho_{S,E}^{(+)}\bigr]$ and $\Theta^\dagger \, \tilde{\rho}_S(\tau/2) \, \Theta = \text{Tr}_E\bigl[ \rho_{S,E}^{(-)}\bigr]$.

The visibility, determined by the off-diagonal elements of the auxiliary degree of freedom, reads in this case:
\begin{align}\label{eqn:visibility2}
\mathcal{V} = &~\Bigl| \mathrm{Tr}_{S,E}\Bigl[ \bigl(U(\tau/2 , 0) \otimes \mathbb{1}_E)\ket{\psi^{(0)}}_{S,E}\bra{\tilde{\psi}^{(0)}}_{S,E} \nonumber \\ 
&~~~ \bigl(\tilde{U}^\dagger(\tau/2 , 0)~\Theta \otimes \mathbb{1}_E\bigr)\Bigr] \Bigr| \nonumber \\
= &~ \Bigl|\bra{\tilde{\psi}^{(0)}}_{S,E}  \bigl(U(\tau , 0) \otimes \mathbb{1}_E\bigr) \ket{\psi^{(0)}}_{S,E} \Bigr|,
\end{align}
which can no longer be related to the different outcomes of a TPM scheme. This notwithstanding, as we will shortly see, one can still make use of this information in an alternative way.

From Ref.~\cite{PhysRevLett.77.2154}, we know that the visibility  $\mathcal{V}$ of the interferometer fringes and the distinguishability $D({\rho},{\sigma})$ between two ``which-path detector states'' ${\rho}$ and ${\sigma}$ (\textit{i.e}., two states from which we can optimally infer the which-path information, would we perform a measurement to distinguish between them) are mutually exclusive. In particular, it has been shown that these two quantities respect the \textit{complementarity relationship} 
\begin{equation}\label{eqn:inter}
\mathcal{V}^2 + D^2({\rho},{\sigma}) \leqslant 1,
\end{equation}
and that this relation becomes an equality if the ``detector states'' are in pure states, as it is in our case.
The distinguishability between the two states is given by the \textit{trace-norm distance} between them, i.e., 
$D({\rho},{\sigma}):= \frac{1}{2} ||{\rho}-{\sigma}|| := \frac{1}{2} \text{Tr} \big[\sqrt{({\rho}-{\sigma})^{\dagger} \, ({\rho}-{\sigma})}\big]$. 

In our case, $D\bigl(\rho_{S,E}^{(+)},\rho_{S,E}^{(-)}\bigr)$ gives us an estimation of how well one can distinguish between the two paths in the interferometer by measuring the system and the environment. However, we are interested in the trace-norm distance between the marginal states of the system only. We can therefore use the fact that the trace distance is non-increasing under partial trace, i.e.,  $D\bigl(\rho_{S,E}^{(+)},\rho_{S,E}^{(-)}\bigr) \geq D\bigl(\rho_{S}(\tau/2),\Theta^\dagger \, \tilde{\rho}_{S}(\tau/2) \, \Theta \bigr)$, to get:
\begin{align} \label{eqn:distrace}
\mathcal{V}^2 + D^2\bigl(\rho_{S}(\tau/2),\Theta^\dagger \, &\tilde{\rho}_{S}(\tau/2) \, \Theta \bigr) \notag \\
& \leqslant \mathcal{V}^2 + D^2\bigl(\rho_{S,E}^{(+)},\rho_{S,E}^{(-)}\bigr) = 1.
\end{align}

Finally, we relate the distinguishability between the system states at $\tau/2$ in the forward and time-reversal dynamics with the relative entropy in Eq.~\eqref{eqn:basic}, and hence to the average dissipative work during the protocol $\Lambda$. This can be done using the upper bounds obtained in Eqs.~(17) and (19) of Ref. \cite{doi:10.1063/1.2044667}. Minor manipulations of these equations lead to the formulation of the following theorem:
\begin{theorem}
Let ${\rho}$ and ${\sigma}$ be two strictly positive density operators in a finite-dimensional Hilbert space $\mathcal{H}$. Then
\begin{equation}
\label{eqn:quadbound}
S({\rho}||{\sigma}) \leqslant \dfrac{||{\rho}-{\sigma}||_2^2}{\alpha_{\sigma}} \leqslant \dfrac{||{\rho}-{\sigma}||^2}{\alpha_{\sigma}}
\end{equation}
where $\alpha_{\sigma} \in (0,1]$ is the smallest eigenvalue of ${\sigma}$, and $||{\varrho}||_2 = \sqrt{\mathrm{Tr}[\varrho^\dagger \, \varrho]}$ denotes the Frobenius (or Euclidean) norm, which verifies $||{\varrho}||_2 \leq ||{\varrho}||$.

Furthermore, setting the dimension of the Hilbert space to $\dim(\mathcal{H}) \equiv d$, we also have:
\begin{align}
\label{eqn:logbound}
S({\rho}||{\sigma}) &\leqslant ||{\rho}-{\sigma}|| \, \mathrm{log}(d/\sqrt{\alpha_{\sigma})} + e^{-1} \notag \\
&= D({\rho},{\sigma}) \, \mathrm{log}(d^2/\alpha_{\sigma}) + e^{-1}.
\end{align}
\end{theorem}

Combining Eqs.~\eqref{eqn:distrace} and the bounds \eqref{eqn:quadbound}-\eqref{eqn:logbound}, we obtain the following two bounds for the dissipative work during the original thermodynamic process:
\begin{align}
\label{eqn:diss_bound1}
\langle W_\mathrm{diss} \rangle &\leqslant k_B T~ 4 \bigl(1 - \mathcal{V}^2\bigr)/\alpha \equiv \mathcal{B}_2, \\
\langle W_\mathrm{diss} \rangle &\leqslant k_B T~ \bigl[\sqrt{1 - \mathcal{V}^2} \,\log(d^2/\alpha) + e^{-1}\bigr] \equiv \mathcal{B}_{\log}, \label{eqn:diss_bound4}
\end{align}
where we have used the relation between the dissipative work and the relative entropy in Eq.~(\ref{eqn:basic}). Additionally, we denoted $\alpha \equiv \alpha_{\tilde{\rho}_S(\tau/2)} = \alpha_{{\rho}_{\tau}^{\text{th}}} = e^{-\beta (E_\mathrm{max}^{\tau} - F_\tau)}$, where $E_{\mathrm{max}}^{\tau}$ is the maximum eigenvalue of the Hamiltonian $H_\tau$. 
This follows from the fact that the states $\tilde{\rho}_S(\tau/2)$ and $\tilde{\rho}_S(0)= \rho_\tau^\text{th}$ have the same spectrum due to their \textit{unitary equivalence}, that is, $\tilde{\rho}_S(\tau/2) = \tilde{U}(\tau/2, 0) \, \tilde{\rho}_S(0) \, \tilde{U}^{\dagger}(\tau/2,0)$.

We notice that the bounds \eqref{eqn:diss_bound1}-\eqref{eqn:diss_bound4} cannot be saturated in general when the initial state of the system is mixed due to the complementarity relation in Eq.~\eqref{eqn:distrace}, which involves a partial trace over the environmental degrees of freedom. (Conversely, saturation would require either measuring the whole environment or a pure initial state of the system as in the previous sections.) Nevertheless, there is a single case where the bound $\mathcal{B}_2$ in Eq.~\eqref{eqn:diss_bound1} is saturated, namely, by verifying the reversibility conditions (quasi-static evolution) where $\mathcal{V} \rightarrow 1$ and $\langle W_\mathrm{diss} \rangle \rightarrow 0$. On the other hand, the bound in Eq.~\eqref{eqn:diss_bound4} is not saturated even in the reversible case, since it is designed to work better in irreversible conditions for $\mathcal{V} < 1$ and $\alpha \rightarrow 0$, where Eq.~\eqref{eqn:distrace} becomes a strict inequality.

Further practical limitations on the ability to split the protocol or to implement the time reversal operation $\Theta$ are addressed in appendix~\ref{Appendix:A}. 

Although in our discussion we supposed that the auxiliary degree-of-freedom is the path of the particle which encodes the system of interest, this is not a requirement of our proposal.
The only three requirements on the auxiliary degree-of-freedom are the following.~(1) The state in Eq.~\eqref{eqn:psi0} should be initially prepared.~(2) Depending on the state of this auxiliary degree-of-freedom, the forward and time-reversal evolutions should then be implemented.~(3) Finally, the auxiliary degree-of-freedom should be measured in the basis $\{\frac{1}{\sqrt{2}}\ket{0}_A\pm \frac{1}{\sqrt{2}}e^{i\phi}\ket{1}_A\}$, while scanning the phase $\phi$ to estimate the visibility.
This auxiliary degree-of-freedom could be encoded in the same particle, perhaps in additional energy levels of an atomic system, in which case the visibility measurement would take the form of atomic interferometry.
Alternatively, a second particle could be used to condition the forward and time-reversal evolution.
For example, if the target system is a qubit encoded in a single trapped ion, one could place a second ion in the trap and then couple the two via the collective vibrational mode.
More concretely,  Ref. \cite{friis2014implementing} shows explicitly how one can implement the controlled evolution of different unitary operations using trapped ions.
In this case, the initial state [Eq.~\eqref{eqn:psi0}] would require entangling operations for its preparation, and the visibility could be easily measured on the internal degree-of-freedom of the second ion.


\section{EXAMPLE OF A PHOTONIC IMPLEMENTATION}

\begin{figure*}[th!]
\centering
\includegraphics[width=0.9\textwidth]{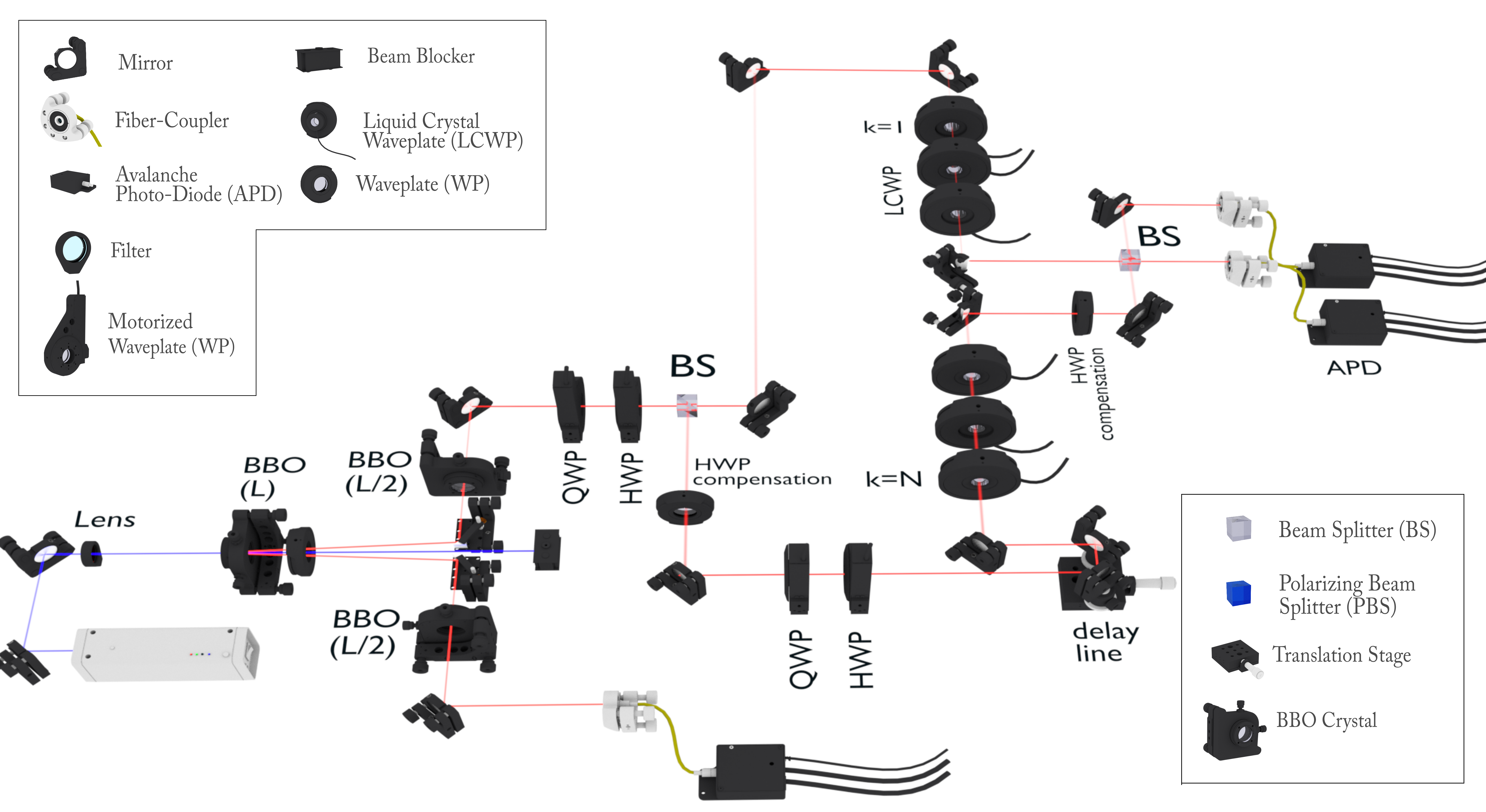}
\caption{Proposed experimental set-up for the photonic implementation of the scheme. Single photon pairs are produced via a type-II spontaneous parametric down conversion (SPDC) source. Photons in each pair are indistinguishable and entangled in polarization. To give rise to pure (mixed) states, one of the two photons is detected with (without) polarization resolution. The remaining single photon is sent through the set-up to realize the quantum measurement scheme. In one arm of the Mach-Zehnder interferometer, the state $\vert E_n^{(0)}\rangle$ is prepared in polarization (via a quarter- (QWP) and a half- (HWP) waveplates), the unitary $U(\tau/2,0)$ is then applied to such a state by means of a sequence of liquid crystal waveplates (LCWPs). In the other arm of the interferometer, the photon is prepared in the state $\vert E_m^{(\tau)}\rangle$ and it is then subjected to the unitary $\Theta^\dagger \, \tilde{U}(\tau/2,0)$. After the two paths are recombined on a beam splitter (BS), the interference fringes are measured by varying the length of the trombone delay-line positioned along one of the two interferometric paths. 
By preparing the initial state of the photon from an entangled state in polarisation, the present scheme can be adapted to operate with initial thermal states (as in Sec.~\ref{sec:lim_prep}).}
\label{img:setup_ThermoIrrev}
\end{figure*}

We apply our scheme to an illustrative experimental set-up in which the thermodynamic system is represented by a single qubit realized through the polarization degree of freedom of a single photon, its thermality is given by the degree of entanglement with an additional photon, the auxiliary qubit is encoded in its path, and the time-dependent thermodynamic process is performed in $N$ discrete time-steps $t_k$ by
sending the photon through a sequence of liquid crystal waveplates each executing a quench on the 
(time-independent) Hamiltonian $H\bigl(\Lambda(t_k)\bigr)$ 
with $k= 1,...,N$, as sketched in Fig.~\ref{img:setup_ThermoIrrev}.

The Hamiltonian of the qubit system can be defined 
as: 
\begin{equation}
\label{eqn:Hamiltonian}
H\bigl(\Lambda\bigr) = \frac{ \hbar \omega}{2} \Bigl[\mathbb{1} + \mathrm{cos}\bigl(\Lambda\bigr) \, {\sigma}_z + \mathrm{sin}\bigl(\Lambda\bigr) \, {\sigma}_x\Bigr],
\end{equation}
where $\omega$ is the qubit's natural frequency, and the control parameter implements $N$ sudden changes 
in the range $\Lambda(0)=0$ to $\Lambda(\tau)= \frac{\pi}{2}$. Consequently, the Hamiltonian is 
given by the spin operator within the $x-z$ plane, which rotates by 
an angle of $\frac{\pi}{2N}$ at each step
around the $y$-axis. At the initial and final times of the protocol, the Hamiltonian is diagonal in the $\sigma_z$ and $\sigma_x$ bases, respectively. Therefore, $\ket{E_n^{(0)}} = \{\ket{z_-} , \ket{z_+}\}$ with corresponding energies $E_n^{(0)}=\{ 0 , \hbar \omega \}$, and $\ket{E_m^{(\tau)}} = \{\ket{x_-}, \ket{x_+}\}$, where  $\ket{x_-}=1/\sqrt{2}(|z_-\rangle -|z_+\rangle)$ and $\ket{x_+}=1/\sqrt{2}(|z_-\rangle +|z_+\rangle))$, with same eigenvalues 
$E_m^{(\tau)}=\{ 0 , \hbar \omega \}$. This implies that $F_0 = F_\tau= -\log(1+e^{-\beta \hbar \omega})$, and thus $\Delta F = F_\tau - F_0 = 0$ such that $W_{\text{diss}}=W$.


In the $k$-th step, the control parameter takes a fixed value 
$\Lambda_k \equiv \Omega \tau k/N$, where $\Omega =\frac{\pi}{2\tau}$ is the angular frequency of the rotation. 
Therefore, any initial state $\ket{\psi_R(0)}$ evolves according to
\begin{equation} \label{eq:sequence}
e^{-\frac{i}{\hbar}H_N \, \Delta t} \cdots e^{-\frac{i}{\hbar}H_2 \, \Delta t} e^{-\frac{i}{\hbar}H_1 \, \Delta t} \ket{\psi_R (0)},
\end{equation}
where $\Delta t = \frac{\pi}{2 N \Omega}$, and for each step $k=1,..., N$:
\begin{equation}
{H}_k = \frac{\hbar \omega}{2} \biggl[\mathbb{1} + \mathrm{cos}\Bigl(\dfrac{k \pi}{2 N}\Bigr) \, {\sigma}_z + \mathrm{sin}\Bigl(\dfrac{k \pi}{2 N}\Bigr) \, {\sigma}_x\biggr].
\end{equation}
The Hamiltonian at each step $H_k$ induces a rotation on the system state of an angle $\theta = \frac{\pi}{2N} \frac{\omega}{\Omega}$ around the axis whose direction $\vec{d}_k = (\sin\bigl(\frac{k\pi}{2N}\bigr), 0, \cos\bigl(\frac{k\pi}{2N}\bigr))$ changes from step to step.
This evolution can be implemented by means of a sequence of $N$ liquid crystal wave-plates (LCWPs). The \textit{k}-th LCWP rotates the photon's polarization about an axis $\vec{d}_k$, and the angle of rotation is given by the retardance which we can change by an externally applied voltage. Hence, to implement the full evolution we can use a series of $N$ LCWPs, each with an optic axis set at $\vartheta_k = \frac{k \pi}{4N} \in [0, \pi/2]$, and with the same retardance for all LCWPs (i.e., $\theta$).

\begin{figure*}[tbh]
\centering
\includegraphics[width=\textwidth]{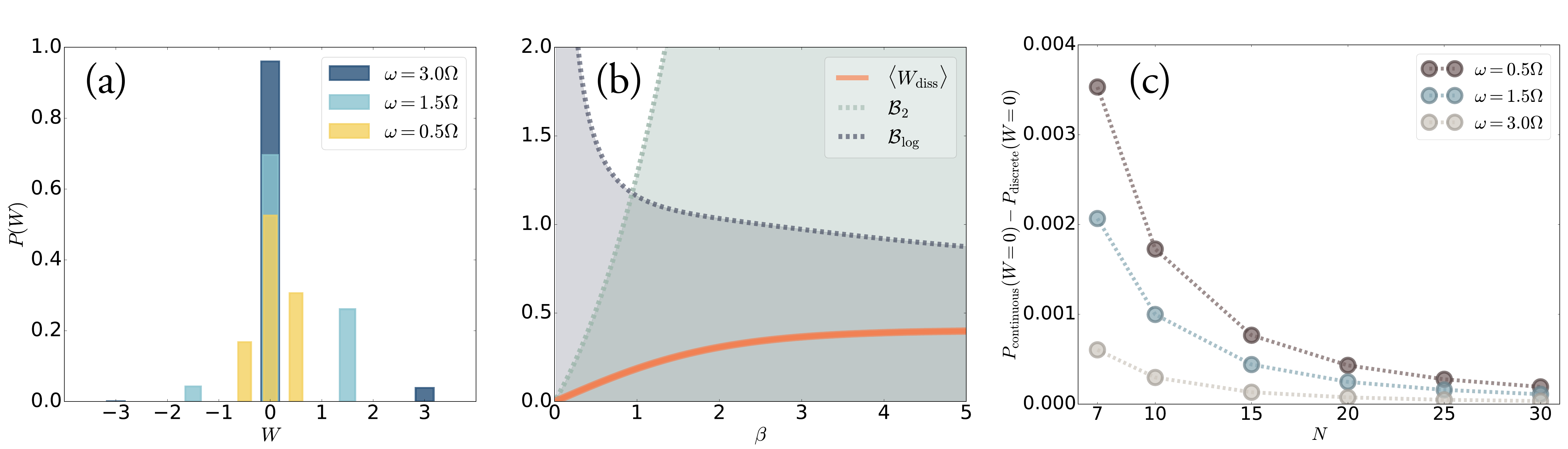}
\caption{Work probability distribution and bounds on the dissipative work for the qubit system under the action of the discrete rotation protocol. (a) Work probability distribution in Eq.~\eqref{eqn:wprob} for three different protocol velocities corresponding to $\omega =\{ 0.5, 1.5, 3.0\} \Omega$ for a fixed inverse temperature $\beta = 1.2 (\hbar \Omega)^{-1}$.
(b) Average dissipative work $\langle W_\mathrm{diss} \rangle$ as defined in Eq.~\eqref{eqn:basic} (orange solid line) and bounds $\mathcal{B}_2$ in Eq.~\eqref{eqn:diss_bound1} (green dashed line) and $\mathcal{B}_\mathrm{log}$ in  Eq.~\eqref{eqn:diss_bound4} (blue dashed line) as a function of the inverse temperature $\beta$ for the case $\omega = \Omega$. Work is expressed in units of $\hbar \omega$.
(c) Difference in the probability $P(W=0)$ between the continuous rotation protocol and the discrete one with $N$ steps as a function of $N$. The three different set of points correspond to the  protocol velocities (grey, blue, purple) corresponding to $\omega =\{ 0.5, 1.5, 3.0\} \Omega$ and $\beta = 1.2 (\hbar \Omega)^{-1}$. In all cases, the difference in probability distributions between the discrete and continuous systems is shown to decrease as the number of steps increases. In this sense, the quench generated by a discrete series of time-independent Hamiltonians is an approximation to the quench generated by a continuous time-dependent Hamiltonian. }
\label{img:bounds}
\end{figure*}

Our scheme can be executed, following Fig.~\ref{img:setup_ThermoIrrev}, by inserting pairs of the  eigenstates of the Hamiltonians $H_0$ and $H_N$ to the interferometer. In particular, we take $N=7$, and apply the discretised Hamiltonian $H_k$ 
for $k = 1, 2, 3$ along path $\ket{0}_A$, while along path $\ket{1}_A$ we perform $H_k$ 
for $k = 6, 5, 4$. In this case, we recover the whole work probability distribution, together with the average dissipative work during the process, which 
can be used to test the fluctuation relations. We note that the work evaluated does not correspond to the intrinsic photonic energy (which is given by its frequency) but to the generator of the evolution~(\ref{eqn:Hamiltonian}). 

In addition, our scheme can also be used to test the upper bounds on the dissipative work obtained in Eqs.~\eqref{eqn:diss_bound1} and \eqref{eqn:diss_bound4} by inserting the thermal states of the two Hamiltonians. More precisely, we may insert a single photon from a pair of photons in a partially entangled state $\ket{\psi}_0^{\text{th}} = a \ket{z_+}\ket{z_+} + b \ket{z_-} \ket{z_-}$, where $a,b\in \mathbb{C}$.
The state of the injected photon is obtained by tracing out the second photon, ${\rho}_0^{\text{th}} = \vert a\vert^2 \ket{z_+}\bra{z_+} + \vert b\vert^2 \ket{z_-}\bra{z_-}$. This corresponds to a thermal state for the choice $\vert a\vert^2 = \frac{\exp(-\beta \hbar \omega)}{Z_0}$ and $\vert b\vert^2= \frac{1}{Z_0}$, with $Z_0 = 1+\exp(-\beta \hbar \omega)$.
Specifically, if $T \rightarrow 0$, $\rho=\ket{z_-}\bra{z_-}$, while if $T \rightarrow \infty$, $\rho=\mathbb{1}/2 $. 


In Fig.~\ref{img:bounds}(a), we show the 
expected work probability distribution associated to our discretized  protocol with $N=7$ steps for a fixed inverse temperature $\beta = 1.2 (\hbar \Omega)^{-1}$ and three values of the frequency $\omega = \{0.5 , 1.5, 3.0 \} \Omega$ 
(light blue, dark blue, yellow) and fixed duration $\tau= 2 \pi/\Omega$. Since the eigenvalues of the Hamiltonian $H\bigl(\Lambda\bigr)$ are constant, the work probability distribution consists of three peaks placed at work values 
$W = \{-\hbar \omega, 0, \hbar \omega \}$. Low temperatures favor an asymmetric distribution with a higher peak for positive work $W=\hbar \omega$ with respect to $W=-\hbar \omega$, while for higher temperatures the two lateral peaks approach equal heights. Moreover, as we observe, for faster protocols the three peaks in the distribution are comparable, while slower protocols $\omega \gg \Omega$ lead to the suppression of lateral peaks in favor of a high central peak at $W=0$. 
In this case, we approach an adiabatic evolution where the initial populations of Hamiltonian eigenstates remain almost constant in time, hence leading to zero energy changes and zero work. On the contrary, in the opposite limit $\omega \ll \Omega$, we approach a sudden quench of the Hamiltonian. In this case, the state of the system remains unchanged by the evolution in Eq.~\eqref{eq:sequence}, and the lateral peaks associated with the overlaps $|\langle x_- | z_+ \rangle|$ and $|\langle x_+ | z_- \rangle|$ become maximal.   


In Fig.~\ref{img:bounds}(b) we show the performance of the bounds for the dissipative work in Eqs.~\eqref{eqn:diss_bound1} and \eqref{eqn:diss_bound4}. We assume the equality in Eq.~\eqref{eqn:distrace}, and take a rotation frequency $\Omega= 1.5 \omega$. As can be appreciated in the plot, in the low temperature regime (right side) the logarithmic bound $\mathcal{B}_{\log}$ becomes the best option while $\mathcal{B}_{2}$ diverges due to the exponential decrease of $\alpha_{\rho_\tau^{\mathrm{th}}} = e^{-\beta [E_\mathrm{max}^\tau - F(\tau)]}$ with temperature. On the contrary, when temperature is increased (left part), $\mathcal{B}_2$ starts to perform better as soon as $k_B T$ becomes higher than the system energy splitting ($k_B T > \hbar \omega$). When increasing $\Omega$ (not shown in the Figure), logarithmic and quadratic bounds become tighter in their respective temperature regimes of performance. 
In the opposite limit of a near adiabatic process (where the dissipative work vanishes), the quadratic bound still performs good for high temperatures, but, contrary to previous cases, the logarithmic bound becomes worst even in the limit of small temperatures. Nevertheless, the bounds do not appear to become saturated in any of the parameters' regime. 

In the limit of many steps $N\gg 1$ 
the discrete rotation protocol can be approximated by a continuous rotation, with $\Lambda(t) = \Omega t$ for arbitrary $\Omega$ and $t \in [0, \tau]$.
Experimentally, this could be realized using ``twisted nematic liquid crystals'' (TNLC). These are devices where the optic axis is continuously rotated (typically by $90^\circ$) along the beam propagation \cite{gooch1975optical}. For our proposal, we would require two devices with $45^\circ$ rotation, one for the forward arm and one for the time-reversed arm.
Note that one could directly implement the final unitary operation using a set of three waveplates [which can implement an arbitrary SU(2) operation]. However, this is not a faithful implementation of the time dependent Hamiltonian as the polarization state will not evolve correctly as it traverses these waveplates.
In such a case, the unitary evolution reads (see App.~\ref{Appendix:B} for details): 
\begin{equation}
U(t,0) = e^{-i \frac{\Omega t}{2} \sigma_y} e^{-\frac{i}{2} \bigl[ \omega \, (\mathbb{1} + \sigma_z)  - \Omega \sigma_y \bigr] \, t}.
\end{equation} 
The differences in the work probability distribution between the discrete and continuous versions of the protocol decreases as the number of steps $N$ increases [Fig.~\ref{img:bounds}(c)].
As before, the ratio $\omega/\Omega$ determines the adiabaticity of the realized process.
Using TNLCs, the length of the liquid crystal cell sets $\Omega$ \cite{gooch1975optical}.
For optical wavelengths, standard TNLCs operate in the our adiabatic regime, using long enough cells with a length of $\approx 10\mu$m. Reaching the non-adiabiatic regime would require cells that are shorter than 2$\mu$m, which should also be achievable with current technology \cite{takatoh2012fast}.
In the limit $\omega/\Omega \gg 1$, we obtain a fully adiabatic process, where the populations of Hamiltonian eigenstates remain constant through the entire evolution (see App.~\ref{Appendix:B} for a detailed analysis). Moreover, since the Hamiltonian $H\bigl(\Lambda(t)\bigr)$ has the same  eigenvalues at all times, we conclude 
that, under adiabatic evolution, a system starting in a thermal state at $t=0$ will remain in equilibrium at the same temperature at all later times. 


\section{CONCLUSIONS}

In this work, we have developed a new method based on interferometric tools to measure the work probability distribution and the thermodynamic irreversibility of a generic driving process acting on a quantum system. The method utilizes the intereference between two paths, one along which the system is driven out of thermal equilibrium in the forward, and one where it is driven in the time-reversal process. We demonstrated that inserting the energy eigenstates of the initial and final Hamiltonians of the system in the two paths of the interferometer and measuring the fringe visibility enable us to directly reconstruct the work distribution and the average dissipative work. The latter is known to be equal to a production of positive average entropy, and it is a measure of the thermodynamic irreversibility.

Our proposal offers a faster implementation speed than TPM schemes as it halves the duration of each execution. A speed enhancement in each run is a considerable advantage since, in TPM schemes, sufficient statistics must be acquired to reconstruct the order of $N^2$ instances (i.e., probabilities) of the work probability distribution from the results of projective measurements. Furthermore, in the TPM scheme the results of the projective measurements are randomly sampled. Due to finite size effects in sampling, the TPM scheme can have a significant delay in acquiring sufficient data, especially for low-probability instances in the work distribution. In contrast, in our scheme one can control which instance in the work distribution to measure by choosing the appropriate input states in the forward and time-reversal amplitudes, making the scheme much less affected by finite-size statistics. Our scheme also offers advantages over existing alternatives to the TPM scheme~\cite{PhysRevLett.110.230602, Serra2014, Serra2015} as it enables a direct measure of the conditional probabilities that make up the work probability distribution. For example, in Refs.~\cite{PhysRevLett.110.230602, Serra2014, Serra2015}, the proposed scheme measures the characteristic function of work, i.e., the Fourier transform of the work probability distribution, from which the work probability distribution must then be recovered indirectly. In the implementations of Refs.~\cite{Serra2014, Serra2015}, this problem required a large sampling of a \textit{continuous} function (the characteristic function) to recover a discrete probability distribution with only a few peaks. In our proposal, these drawbacks are overcome by directly obtaining the conditional probabilities associated with the peaks.

In the case of limited experimental control, when only the thermal states of the initial and final Hamiltonians of the system can be prepared, our method provides useful upper bounds on the average dissipative work. The scheme involves no entangling operations with external auxiliary systems and no energy measurements, and thus offers an accessible and versatile playground for studying the thermodynamics of quantum processes. 

To provide a concrete example of implementation of our scheme, we have 
developed an experimental proposal of our scheme using an all-optical platform, and standard tools for single- and entangled-photon manipulation. The out-of-equilibrium quantum dynamics is realized via a series of liquid crystals wave-plates splitting the thermodynamic process in a series of discrete time-steps $t_k$, each represented by a liquid crystal with an optical axis set at a different angle of rotation $\vartheta_k$. 

Although here we have focused for simplicity on the case of initial equilibrium states, we stress that our method can be used to determine the work probability distribution for generic nonequilibrium initial states. Systems with initial coherence in the energy basis or composite systems sharing quantum correlations can also be handled within our method by measuring transitions from arbitrary eigenstates $p_{m|n}=|\langle E_m^{(\tau)}| U(\tau,0)|\psi_n^{(0)}\rangle|^2$ and using extended trajectories (Bayesian networks) techniques~\cite{Micadei_2020,PhysRevE.101.052128} to infer the work probability distribution. Finally, work probability distributions  using collective measurements might be instead reproduced following the proposal in Ref.~\cite{Wueaav4944} by augmenting the number of paths and using them to encode other system degrees of freedom.

\vfill

\renewcommand{\baselinestretch}{1.2}

\begin{acknowledgments}
\vspace{2mm}
\textbf{Acknowledgments:} The authors wish to thank the organisers of the conference ``\textit{New Directions in Quantum Information}'', (Nordita, Stockholm (Sweden); April 1-26 2019) for providing a stimulating platform for the discussion which originated this result. \textbf{Funding:} G.R. acknowledges financial support from the Royal Society through the Newton International
Fellowship No.\ NIF$\backslash \text{R1}\backslash$202512. G.M. acknowledges funding from Spanish MICINN through the Juan de la Cierva program (IJC2019-039592-I) and the European Union's Horizon 2020 research and innovation program under the Marie Sk\l{}odowska-Curie grant agreement No 801110 and the Austrian Federal Ministry of Education, Science and Research (BMBWF). L.A.R. acknowledges financial support from the Austrian Science Fund (FWF) through BeyondC (F7113). P.W. acknowledges financial support from the research platform TURIS, the  European Commission through EPIQUS (no. 899368), and from the Austrian Science Fund via GRIPS(P30817-N36), BeyondC (F7113) and Research Group 5 (FG5). J.M.R.P. acknowledges financial support from the Spanish Government (Grant Contract, FIS-2017-83706-R). \v C.B. acknowledges financial support from the  Austrian  Science  Fund  (FWF)  through  the SFB project BeyondC (sub-project F7103), a grant from the Foundational Questions Institute (FQXi) Fund, as well as from the European Commission via Testing  the  Large-Scale  Limit  of  Quantum  Mechanics (TEQ) (No. 766900) project. This publication was made possible through the support of the ID61466 grant from the John Templeton Foundation, as part of the The Quantum Information Structureof Spacetime (QISS) Project (qiss.fr). The opinions expressed in this publication are those of the authors and do not necessarily reflect the views of the John Templeton Foundation.
\end{acknowledgments}


\appendix

\section{Limited control over experimental settings} \label{Appendix:A}

In this section, we consider the situation where the ability to control the application of the protocol $\Lambda$ is heavily affected by experimental limitations such as \textit{(i)}. impossibility to split the protocol $\Lambda$ in two halves and invert the second half,  
or \textit{(ii).} difficulties in applying the time-reversal operation $\Theta^\dagger$ at the end of the second branch of the interferometer. If any of these circumstances applies, the requirements for the usability of the interferometric scheme proposed above may not be met. In light of this, here we propose an alternative set-up to be applied in such situations. The main price to pay is that the time needed to run the scheme for any initial state is doubled.

In this alternative scheme, we will take advantage of the unitary equivalence of the system states in the forward and time-reversal dynamics. In addition, the relation between the dissipative work and the relative entropy in Eq.~\eqref{eqn:basic} is verified for any intermediate instant of time $t \in [0, \tau]$. As a consequence, we can observe interference between the states in the forward and time-reverse dynamics also at the extremes of the interval, where one of the two states is thermal. In the following, we present the scheme in the case of interference at time $\tau$ in the forward dynamics (corresponding to $t=0$ in the time-reversal dynamics), but an analogous scheme can be developed for interference at time $t=0$ in the forward dynamics (corresponding to $t=\tau$ in the time-reversal one).

As in the previous case, we start by preparing the auxiliary degree of freedom in the quantum superposition $\frac{1}{\sqrt{2}}\bigl(\ket{0}_{A} + \ket{1}_{A} \bigr)$ at $t<0$. Once again, the initial states of the system in the two branches may either be the pure states $\ket{E_n^{(0)}}$ along the path $\ket{0}_A$ and $ \ket{E_m^{(\tau)}}$ along $\ket{1}_A$, or the mixed thermal states $\rho_0^\mathrm{th}$ and $\rho_\tau^\mathrm{th}$, respectively, depending on whether we have full control over the system in the preparation stage. However, in contrast to the previous case, we implement the whole protocol $\Lambda$ over the system in the path $\ket{0}_A$, while the branch $\ket{1}_A$ remains unaffected.

Assuming, for concreteness, initial pure states, the global state of the system and the auxiliary system after time $\tau$ can be evaluated and, tracing the system degrees of freedom, we obtain:
\begin{align} \label{eqn:rho-ancilla-new}
&\rho_{A}(\tau):= \text{Tr}_{S}[\rho_{S,A}(\tau)] = \dfrac{1}{2}\Bigl\{\ket{0}_{A} \bra{0}_{A} +~ \ket{1}_{A} \bra{1}_{A} \notag \\
&+ \ket{0}_{A} \bra{1}_{A} \text{Tr}_{S} \Bigl[U(\tau,0)  \ket{E_n^{(0)}} \bra{E_m^{(\tau)}} \Bigr] +  \notag\\
&+ \ket{1}_{A} \bra{0}_{A} \text{Tr}_{S} \Bigl[ \ket{E_m^{(\tau)}}\bra{E_n^{(0)}} U^\dagger(\tau,0) \Bigr]\Bigr\}.
\end{align}
Consequently, in this case the visibility directly give us the conditional probabilities for the work probability distribution: 
\begin{equation}\label{eqn:visibility1}
\mathcal{V}_{m, n}:=\Bigl|\text{Tr}_{S} \Bigl[U(\tau,0)  \ket{E_n^{(0)}} \bra{E_m^{(\tau)}} \Bigr] \Bigr|= \sqrt{p_{m|n}},
\end{equation}
and we recover Eqs.~\eqref{eqn:workprob} and \eqref{eqn:avwork}.

Likewise, when the initial states in the two interferometer paths are the mixed thermal states, we find again, for the visibility: 
\begin{align}\label{eqn:visibility3}
\mathcal{V} = &~\Bigl| \mathrm{Tr}_{S,E}[ (U(\tau , 0) \otimes \mathbb{1}_E)\ket{\psi^{(0)}}_{S,E}\bra{\tilde{\psi}^{(0)}}_{S,E}] \Bigr| \nonumber \\
= &~ \Bigl|\bra{\tilde{\psi}^{(0)}}_{S,E}  (U(\tau , 0) \otimes \mathbb{1}_E) \ket{\psi^{(0)}}_{S,E} \Bigr|,
\end{align}
which is equivalent to Eq.~\eqref{eqn:visibility2}. Consequently, the bounds developed in Eqs.~\eqref{eqn:diss_bound1} and \eqref{eqn:diss_bound4} for the dissipative work apply also in this situation.

\section{Continuous Rotation Protocol} \label{Appendix:B}

We analytically obtain the evolution generated by the time-dependent Hamiltonian in Eq.~\eqref{eqn:Hamiltonian} for the continuous variation of the control parameter $\Lambda(t) = \Omega t$ for a constant angular velocity $\Omega$ in the interval $t \in [0,\tau]$.
In order to reach a description in terms of a time-independent Hamiltonian, we use a picture in which the states rotate at the same rate as the Hamiltonian around the $\vec{y}$-axis: $\ket{\psi(t)} = e^{-i\frac{\Omega t}{2} \sigma_y} \ket{\psi_0(t)}$.

We write the time-dependent Schr\"odinger equation $i \hbar \frac{d}{dt} \ket{\psi(t)} = H\bigl(\Lambda(t)\bigr) \ket{\psi(t)}$ with this substitution for $\ket{\psi(t)}$, as
\begin{align}
\label{eqn:evol}
&\dfrac{\hbar  \Omega}{2}\sigma_y \ket{\psi_0(t)} + i \hbar \dfrac{d}{dt} \ket{\psi_0(t)} = \notag\\
& \qquad \qquad e^{i\frac{\Omega t}{2} \sigma_y} H\bigl(\Lambda(t)\bigr) e^{- i\frac{\Omega t}{2} \sigma_y} \ket{\psi_0(t)}.
\end{align}
We focus now on the r. h. s. of this equation. By substituting the expression Eq. (\ref{eqn:Hamiltonian}) for the Hamiltonian, we get
\begin{align}
\label{eqn:H(t)}
e^{i\frac{\Omega t}{2} \sigma_y}& H\bigl(\Lambda(t)\bigr) e^{- i\frac{\Omega t}{2} \sigma_y} \ket{\psi_0(t)} =\notag\\
&= \dfrac{\hbar \omega}{2} \, \Bigl[ \mathbb{1} + \mathrm{cos}\bigl(\Omega t\bigr) e^{i\frac{\Omega t}{2} \sigma_y} \sigma_z e^{- i\frac{\Omega t}{2} \sigma_y} +\notag\\
&\qquad + \mathrm{sin}\bigl(\Omega t\bigr) e^{i\frac{\Omega t}{2} \sigma_y} \sigma_x e^{- i\frac{\Omega t}{2} \sigma_y} \Bigr] \ket{\psi_0(t)}
\end{align}
We now write the Pauli matrices in the $\sigma_y$ operator's eigenbasis, 
and we correspondingly evaluate the two terms in Eq. (\ref{eqn:H(t)}):
\begin{subequations}
\begin{align}
& \cos{(\Omega t)}\sigma_z -\sin{(\Omega t)} \sigma_x
=
e^{i\frac{\Omega t}{2} \sigma_y} \sigma_z
e^{-i\frac{\Omega t}{2} \sigma_y} \\
& \cos{(\Omega t)}\sigma_x +\sin{(\Omega t)} \sigma_z
=
e^{i\frac{\Omega t}{2} \sigma_y} \sigma_x
e^{-i\frac{\Omega t}{2} \sigma_y}
\end{align}
\end{subequations}
From this, Eq. (\ref{eqn:H(t)}) becomes $e^{i\frac{\Omega t}{2} \sigma_y} H\bigr(\Lambda(t)\bigl) e^{- i\frac{\Omega t}{2} \sigma_y} = \, \dfrac{\hbar \omega}{2} \bigl(\mathbb{1} + \sigma_z\bigr)$. By substituting this result into Eq. (\ref{eqn:evol}), we obtain
\begin{equation}
\label{eqn:evol1}
i \hbar \dfrac{d}{dt} \ket{\psi_0(t)} = \dfrac{\hbar}{2} \Bigl[ \omega \bigl(\mathbb{1} + \sigma_z\bigr)  - \Omega \sigma_y \Bigr] \ket{\psi_0(t)}.
\end{equation}

We have thus reduced the Schr\"odinger equation with a time-dependent Hamiltonian into one with a time-independent Hamiltonian. By calling
\begin{equation}
\label{eqn:sincos}
\mathrm{sin}\xi = \dfrac{\omega}{\sqrt{\omega^2 + \Omega^2}} \quad \qquad \mathrm{cos}\xi = \dfrac{-\Omega}{\sqrt{\omega^2 + \Omega^2}},
\end{equation}
where by $\xi$ we defined the angle between the direction $\vec{n} = (0, \mathrm{cos}\xi, \mathrm{sin}\xi)$  and $\vec{y}$-axis within $z$-$y$ plane, we can therefore rewrite Eq. (\ref{eqn:evol1}) as
\begin{equation}
i \hbar \dfrac{d}{dt} \ket{\psi_0(t)} = \dfrac{\hbar \omega}{2} \mathbb{1} + \dfrac{\hbar}{2} \sqrt{\omega^2 + \Omega^2} \,  \vec{n}\cdot \vec{\sigma} \ket{\psi_0(t)}.
\end{equation}
The solution of this equation is $\ket{\psi_0(t)} = e^{-i \frac{\omega}{2} t -\frac{i}{2}  \sqrt{\omega^2 + \Omega^2} \,  \vec{n}\cdot \vec{\sigma} \, t} \ket{\psi_0(0)}$. Neglecting the global phase $(e^{-i\frac{\omega}{2}t})$, we thus get
\begin{align}
\label{eqn:psi(t)}
\ket{\psi(t)} &= e^{- i\frac{\Omega t}{2} \sigma_y} \ket{\psi_0(t)}\\
&= \mathrm{exp}\Bigl(-i\frac{\Omega t}{2} \, \sigma_y\Bigr) \, \mathrm{exp}\biggl(- \frac{i}{2} \sqrt{\omega^2 + \Omega^2} \,  \vec{n}\cdot \vec{\sigma}\, t\biggr) \ket{\psi(0)}. \notag
\end{align}
We now introduce an eigenstate basis $\{\ket{\vec{n}_{\pm}} \}$ of $\vec{n}\cdot \vec{\sigma}$ (i.e., $\vec{n}\cdot \vec{\sigma} \ket{\vec{n}_{\pm}} = \pm \ket{\vec{n}_{\pm}}$), where $\ket{\vec{n}_{+}}=\mathrm{cos}\bigl(\xi/2\bigr)\ket{y_{+}} + \mathrm{sin}\bigl(\xi/2\bigr)\ket{y_{-}}$, and $\ket{\vec{n}_{-}}=- \mathrm{sin}\bigl(\xi/2\bigr)\ket{y_{+}} + \mathrm{cos}\bigl(\xi/2\bigr)\ket{y_{-}}$, with $\{\ket{y_{\pm}} \}$ being the eigenbasis of $\sigma_y$.

If we write the initial state $\ket{\psi_0(0)}$ in terms of this new basis in the general form $\ket{\psi_0(0)} = c_1 \ket{\vec{n}_{+}} + c_2 \ket{\vec{n}_{-}}$, where $c_1$ and $c_2$ are complex numbers, it is straightforward to obtain a solution for Eq. (\ref{eqn:psi(t)}):
\begin{align}
\ket{\psi(t)} = \Bigl[ &c_1 \, e^{-\frac{i}{2} \sqrt{\omega^2 + \Omega^2}\, t}\,\mathrm{cos}\bigl(\xi/2\bigr) \notag\\
&- c_2 \, e^{\frac{i}{2} \sqrt{\omega^2 + \Omega^2}\, t}\,\mathrm{sin}\bigl(\xi/2\bigr)\Bigr] \, e^{-i \frac{\Omega t}{2}} \ket{y_{+}} \notag\\
+ \Bigl[ &c_1 \, e^{-\frac{i}{2} \sqrt{\omega^2 + \Omega^2}\, t}\,\mathrm{sin}\bigl(\xi/2\bigr) \notag\\
&+ c_2 \, e^{\frac{i}{2} \sqrt{\omega^2 + \Omega^2}\, t}\,\mathrm{cos}\bigl(\xi/2\bigr)\Bigr] \, e^{i \frac{\Omega t}{2}} \ket{y_{-}}.
\end{align}
Finally, to set $c_1$ and $c_2$, we suppose that the initial state $\ket{\psi_0(0)}$ was an eigenstate of the initial Hamiltonian $H(t=0)=\frac{\hbar \omega}{2} (\mathbb{1} + \, {\sigma}_z)$, i.e.,
$\ket{z_{+}}=(\ket{y_{+}} + \ket{y_{-}})/\sqrt{2}$ and 
$\ket{z_{-}}=-i(\ket{y_{+}} - \ket{y_{-}})/\sqrt{2}$. Then $c_1 = \bigl[\mathrm{cos}(\xi/2)+\mathrm{sin}(\xi/2)\bigr]/\sqrt{2}$ and $ c_2 = -\bigl[\mathrm{sin}(\xi/2)-\mathrm{cos}(\xi/2)\bigr]/\sqrt{2}$ for $|z_+\rangle$ and
$c_1 =-i \bigl[\mathrm{cos}(\xi/2)-\mathrm{sin}(\xi/2)\bigr]/\sqrt{2}$ and $ c_2 =i \bigl[\mathrm{sin}(\xi/2)+\mathrm{cos}(\xi/2)\bigr]/\sqrt{2}$ for $|z_-\rangle$.

Let us denote by $\vert\psi_{\pm}(\tau)\rangle$ the final states evolved from the two initial states $\vert z_\pm\rangle$. If initially the system was in the thermal state, i.e.,  $\rho^\text{th}_0 = \frac{1}{Z_0} \bigl[\ket{z_-}\bra{z_-}+ e^{-\hbar \omega \beta} \ket{z_+}\bra{z_+}\bigr]$, then the final state will be 
\begin{equation}
\rho_\tau =  \frac{1}{Z_0} \bigl[\ket{\psi_-(\tau)}\bra{\psi_-(\tau)}+ e^{-\hbar \omega \beta} \ket{\psi_+(\tau)}\bra{\psi_+(\tau)}\bigr].
\label{eqn:rho-tau}
\end{equation}


Let us separate the two cases of a slow-varying Hamiltonian $\omega\gg \Omega$ (which will correspond to an adiabatic thermodynamic process), and that of a rapidly-varying Hamiltonian $\omega\ll \Omega$. We will assume below that the initial state is always $|z+\rangle$.

\paragraph{Slow-Varying Hamiltonian ($\omega\gg \Omega$).} Recalling the definitions given in Eq. (\ref{eqn:sincos}), we obtain $\mathrm{sin}\xi \simeq 1$ and $\mathrm{cos}\xi \simeq 0$, and hence $\xi \simeq \pi/2$. In this case $\ket{\psi(t)}_{\omega\gg \Omega} \simeq \bigl(e^{-i \frac{\Omega t}{2}} \ket{y_{+}} + e^{i \frac{\Omega t}{2}} \ket{y_{-}}\bigr)/\sqrt{2}$,
where we have neglected the global phase $e^{-i\omega t}$. We recall that, in the $\sigma_y$ eigenbasis, $\sigma_x \ket{y_{\pm}} = \pm i \ket{y_{\mp}}$ and $\sigma_z \ket{y_{\pm}} = \ket{y_{\mp}}$, the action of the Hamiltonian $H\bigl(\Lambda(t)\bigr)$ onto the state $\ket{\psi(t)}$ in case of an adiabatic evolution of the system will result in
$H\bigl(\Lambda(t)\bigr)\ket{\psi(t)}_{\omega\gg \Omega} \simeq \hbar \omega \ket{\psi(t)}_{\omega \gg \Omega}$. We conclude that, if a system is in an eigenstate of the time-dependent Hamiltonian (\ref{eqn:Hamiltonian}) at time $t=0$, it remains in the eigenstate of the Hamiltonian at any later time $t$. Since the Hamiltonian $H\bigl(\Lambda(t)\bigr)$ has the same  eigenvalues at all times, we conclude from Eq. (\ref{eqn:rho-tau}) that, under the adiabatic evolution, a system starting in a thermal state of the Hamiltonian (\ref{eqn:Hamiltonian}) at $t=0$ at temperature $T$ will remain in the thermal state of the Hamiltonian at the temperature $T$ at all later times. Specifically, the system will end up in the thermal state of $\sigma_x$ at $t=\tau$. In this sense, the system ``thermalizes'' under a slow change of the Hamiltonian. 

\paragraph{Rapidly-Varying Hamiltonian ($\omega\ll \Omega$).} In this case, $\mathrm{sin}\xi \simeq 0$ and $\mathrm{cos}\xi \simeq -1$, thus $\xi \simeq \pi$:
\begin{equation}
\ket{\psi(t)}_{\omega\ll \Omega} \simeq \dfrac{\ket{y_{+}}+\ket{y_{-}}}{\sqrt{2}}.
\end{equation}
We therefore conclude that the state does not depend on time, which indicates that, under a rapid change of the Hamiltonian, the system remains in its initial state.

\bibliography{QDC_BIB}

\begin{thebibliography}{58}%
\makeatletter
\providecommand \@ifxundefined [1]{%
 \@ifx{#1\undefined}
}%
\providecommand \@ifnum [1]{%
 \ifnum #1\expandafter \@firstoftwo
 \else \expandafter \@secondoftwo
 \fi
}%
\providecommand \@ifx [1]{%
 \ifx #1\expandafter \@firstoftwo
 \else \expandafter \@secondoftwo
 \fi
}%
\providecommand \natexlab [1]{#1}%
\providecommand \enquote  [1]{``#1''}%
\providecommand \bibnamefont  [1]{#1}%
\providecommand \bibfnamefont [1]{#1}%
\providecommand \citenamefont [1]{#1}%
\providecommand \href@noop [0]{\@secondoftwo}%
\providecommand \href [0]{\begingroup \@sanitize@url \@href}%
\providecommand \@href[1]{\@@startlink{#1}\@@href}%
\providecommand \@@href[1]{\endgroup#1\@@endlink}%
\providecommand \@sanitize@url [0]{\catcode `\\12\catcode `\$12\catcode
  `\&12\catcode `\#12\catcode `\^12\catcode `\_12\catcode `\%12\relax}%
\providecommand \@@startlink[1]{}%
\providecommand \@@endlink[0]{}%
\providecommand \url  [0]{\begingroup\@sanitize@url \@url }%
\providecommand \@url [1]{\endgroup\@href {#1}{\urlprefix }}%
\providecommand \urlprefix  [0]{URL }%
\providecommand \Eprint [0]{\href }%
\providecommand \doibase [0]{https://doi.org/}%
\providecommand \selectlanguage [0]{\@gobble}%
\providecommand \bibinfo  [0]{\@secondoftwo}%
\providecommand \bibfield  [0]{\@secondoftwo}%
\providecommand \translation [1]{[#1]}%
\providecommand \BibitemOpen [0]{}%
\providecommand \bibitemStop [0]{}%
\providecommand \bibitemNoStop [0]{.\EOS\space}%
\providecommand \EOS [0]{\spacefactor3000\relax}%
\providecommand \BibitemShut  [1]{\csname bibitem#1\endcsname}%
\let\auto@bib@innerbib\@empty
\bibitem [{\citenamefont {Eddington}(1928)}]{Eddington_1928}%
  \BibitemOpen
  \bibfield  {author} {\bibinfo {author} {\bibfnamefont {A.~S.}\ \bibnamefont
  {Eddington}},\ }\href@noop {} {\emph {\bibinfo {title} {The nature of the
  physical world}}}\ (\bibinfo  {publisher} {The University Press},\ \bibinfo
  {address} {Cambridge, England},\ \bibinfo {year} {1928})\BibitemShut
  {NoStop}%
\bibitem [{\citenamefont {Callen}(1985)}]{Callen1985}%
  \BibitemOpen
  \bibfield  {author} {\bibinfo {author} {\bibfnamefont {H.~B.}\ \bibnamefont
  {Callen}},\ }\href@noop {} {\emph {\bibinfo {title} {Thermodynamics and an
  Introduction to Thermostatistics}}}\ (\bibinfo  {publisher} {Wiley},\
  \bibinfo {address} {New York},\ \bibinfo {year} {1985})\BibitemShut {NoStop}%
\bibitem [{\citenamefont {Evans}\ \emph {et~al.}(1993)\citenamefont {Evans},
  \citenamefont {Cohen},\ and\ \citenamefont {Morriss}}]{EvansPRL}%
  \BibitemOpen
  \bibfield  {author} {\bibinfo {author} {\bibfnamefont {D.~J.}\ \bibnamefont
  {Evans}}, \bibinfo {author} {\bibfnamefont {E.~G.~D.}\ \bibnamefont
  {Cohen}},\ and\ \bibinfo {author} {\bibfnamefont {G.~P.}\ \bibnamefont
  {Morriss}},\ }\bibfield  {title} {\bibinfo {title} {Probability of second law
  violations in shearing steady states},\ }\href
  {https://doi.org/10.1103/PhysRevLett.71.2401} {\bibfield  {journal} {\bibinfo
   {journal} {Phys. Rev. Lett.}\ }\textbf {\bibinfo {volume} {71}},\ \bibinfo
  {pages} {2401} (\bibinfo {year} {1993})}\BibitemShut {NoStop}%
\bibitem [{\citenamefont {Jarzynski}(1997)}]{JarzynskiPRL}%
  \BibitemOpen
  \bibfield  {author} {\bibinfo {author} {\bibfnamefont {C.}~\bibnamefont
  {Jarzynski}},\ }\bibfield  {title} {\bibinfo {title} {Nonequilibrium equality
  for free energy differences},\ }\href
  {https://doi.org/10.1103/PhysRevLett.78.2690} {\bibfield  {journal} {\bibinfo
   {journal} {Phys. Rev. Lett.}\ }\textbf {\bibinfo {volume} {78}},\ \bibinfo
  {pages} {2690} (\bibinfo {year} {1997})}\BibitemShut {NoStop}%
\bibitem [{\citenamefont {Crooks}(1999)}]{CrooksPRE}%
  \BibitemOpen
  \bibfield  {author} {\bibinfo {author} {\bibfnamefont {G.~E.}\ \bibnamefont
  {Crooks}},\ }\bibfield  {title} {\bibinfo {title} {Entropy production
  fluctuation theorem and the nonequilibrium work relation for free energy
  differences},\ }\href {https://doi.org/10.1103/PhysRevE.60.2721} {\bibfield
  {journal} {\bibinfo  {journal} {Phys. Rev. E}\ }\textbf {\bibinfo {volume}
  {60}},\ \bibinfo {pages} {2721} (\bibinfo {year} {1999})}\BibitemShut
  {NoStop}%
\bibitem [{\citenamefont {Jarzynski}(2011)}]{JarzynskiREV}%
  \BibitemOpen
  \bibfield  {author} {\bibinfo {author} {\bibfnamefont {C.}~\bibnamefont
  {Jarzynski}},\ }\bibfield  {title} {\bibinfo {title} {Equalities and
  inequalities: Irreversibility and the second law of thermodynamics at the
  nanoscale},\ }\href
  {https://doi.org/10.1146/annurev-conmatphys-062910-140506} {\bibfield
  {journal} {\bibinfo  {journal} {Annual Review of Condensed Matter Physics}\
  }\textbf {\bibinfo {volume} {2}},\ \bibinfo {pages} {329–351} (\bibinfo
  {year} {2011})},\ \Eprint
  {https://arxiv.org/abs/https://doi.org/10.1146/annurev-conmatphys-062910-140506}
  {https://doi.org/10.1146/annurev-conmatphys-062910-140506} \BibitemShut
  {NoStop}%
\bibitem [{\citenamefont {Seifert}(2012)}]{SeifertREV}%
  \BibitemOpen
  \bibfield  {author} {\bibinfo {author} {\bibfnamefont {U.}~\bibnamefont
  {Seifert}},\ }\bibfield  {title} {\bibinfo {title} {Stochastic
  thermodynamics, fluctuation theorems and molecular machines},\ }\href@noop {}
  {\bibfield  {journal} {\bibinfo  {journal} {Rep. Prog. Phys.}\ }\textbf
  {\bibinfo {volume} {75}},\ \bibinfo {pages} {126001} (\bibinfo {year}
  {2012})}\BibitemShut {NoStop}%
\bibitem [{\citenamefont {Esposito}\ \emph {et~al.}(2009)\citenamefont
  {Esposito}, \citenamefont {Harbola},\ and\ \citenamefont
  {Mukamel}}]{RevModPhys.81.1665}%
  \BibitemOpen
  \bibfield  {author} {\bibinfo {author} {\bibfnamefont {M.}~\bibnamefont
  {Esposito}}, \bibinfo {author} {\bibfnamefont {U.}~\bibnamefont {Harbola}},\
  and\ \bibinfo {author} {\bibfnamefont {S.}~\bibnamefont {Mukamel}},\
  }\bibfield  {title} {\bibinfo {title} {Nonequilibrium fluctuations,
  fluctuation theorems, and counting statistics in quantum systems},\ }\href
  {https://doi.org/10.1103/RevModPhys.81.1665} {\bibfield  {journal} {\bibinfo
  {journal} {Rev. Mod. Phys.}\ }\textbf {\bibinfo {volume} {81}},\ \bibinfo
  {pages} {1665} (\bibinfo {year} {2009})}\BibitemShut {NoStop}%
\bibitem [{\citenamefont {Campisi}\ \emph {et~al.}(2011)\citenamefont
  {Campisi}, \citenamefont {H\"anggi},\ and\ \citenamefont
  {Talkner}}]{RevModPhys.83.771}%
  \BibitemOpen
  \bibfield  {author} {\bibinfo {author} {\bibfnamefont {M.}~\bibnamefont
  {Campisi}}, \bibinfo {author} {\bibfnamefont {P.}~\bibnamefont {H\"anggi}},\
  and\ \bibinfo {author} {\bibfnamefont {P.}~\bibnamefont {Talkner}},\
  }\bibfield  {title} {\bibinfo {title} {Colloquium: Quantum fluctuation
  relations: Foundations and applications},\ }\href
  {https://doi.org/10.1103/RevModPhys.83.771} {\bibfield  {journal} {\bibinfo
  {journal} {Rev. Mod. Phys.}\ }\textbf {\bibinfo {volume} {83}},\ \bibinfo
  {pages} {771} (\bibinfo {year} {2011})}\BibitemShut {NoStop}%
\bibitem [{\citenamefont {Funo}\ \emph {et~al.}(2018)\citenamefont {Funo},
  \citenamefont {Ueda},\ and\ \citenamefont {Sagawa}}]{Funo2018}%
  \BibitemOpen
  \bibfield  {author} {\bibinfo {author} {\bibfnamefont {K.}~\bibnamefont
  {Funo}}, \bibinfo {author} {\bibfnamefont {M.}~\bibnamefont {Ueda}},\ and\
  \bibinfo {author} {\bibfnamefont {T.}~\bibnamefont {Sagawa}},\ }\bibinfo
  {title} {Quantum fluctuation theorems},\ in\ \href
  {https://doi.org/10.1007/978-3-319-99046-0_10} {\emph {\bibinfo {booktitle}
  {Thermodynamics in the Quantum Regime}}},\ \bibinfo {editor} {edited by\
  \bibinfo {editor} {\bibfnamefont {F.}~\bibnamefont {Binder}}, \bibinfo
  {editor} {\bibfnamefont {L.~A.}\ \bibnamefont {Correa}}, \bibinfo {editor}
  {\bibfnamefont {C.}~\bibnamefont {Gogolin}}, \bibinfo {editor} {\bibfnamefont
  {J.}~\bibnamefont {Anders}},\ and\ \bibinfo {editor} {\bibfnamefont
  {G.}~\bibnamefont {Adesso}}}\ (\bibinfo  {publisher} {Springer International
  Publishing},\ \bibinfo {address} {Cham},\ \bibinfo {year} {2018})\ pp.\
  \bibinfo {pages} {249--273}\BibitemShut {NoStop}%
\bibitem [{\citenamefont {{Chetrite}}\ and\ \citenamefont
  {{Mallick}}(2012)}]{Chetrite:2012}%
  \BibitemOpen
  \bibfield  {author} {\bibinfo {author} {\bibfnamefont {R.}~\bibnamefont
  {{Chetrite}}}\ and\ \bibinfo {author} {\bibfnamefont {K.}~\bibnamefont
  {{Mallick}}},\ }\bibfield  {title} {\bibinfo {title} {{Quantum Fluctuation
  Relations for the Lindblad Master Equation}},\ }\href
  {https://doi.org/10.1007/s10955-012-0557-z} {\bibfield  {journal} {\bibinfo
  {journal} {Journal of Statistical Physics}\ }\textbf {\bibinfo {volume}
  {148}},\ \bibinfo {pages} {480} (\bibinfo {year} {2012})},\ \Eprint
  {https://arxiv.org/abs/1112.1303} {arXiv:1112.1303 [cond-mat.stat-mech]}
  \BibitemShut {NoStop}%
\bibitem [{\citenamefont {Albash}\ \emph {et~al.}(2013)\citenamefont {Albash},
  \citenamefont {Lidar}, \citenamefont {Marvian},\ and\ \citenamefont
  {Zanardi}}]{PhysRevE.88.032146}%
  \BibitemOpen
  \bibfield  {author} {\bibinfo {author} {\bibfnamefont {T.}~\bibnamefont
  {Albash}}, \bibinfo {author} {\bibfnamefont {D.~A.}\ \bibnamefont {Lidar}},
  \bibinfo {author} {\bibfnamefont {M.}~\bibnamefont {Marvian}},\ and\ \bibinfo
  {author} {\bibfnamefont {P.}~\bibnamefont {Zanardi}},\ }\bibfield  {title}
  {\bibinfo {title} {Fluctuation theorems for quantum processes},\ }\href
  {https://doi.org/10.1103/PhysRevE.88.032146} {\bibfield  {journal} {\bibinfo
  {journal} {Phys. Rev. E}\ }\textbf {\bibinfo {volume} {88}},\ \bibinfo
  {pages} {032146} (\bibinfo {year} {2013})}\BibitemShut {NoStop}%
\bibitem [{\citenamefont {Rastegin}\ and\ \citenamefont {\ifmmode~\dot{Z}\else
  \.{Z}\fi{}yczkowski}(2014)}]{PhysRevE.89.012127}%
  \BibitemOpen
  \bibfield  {author} {\bibinfo {author} {\bibfnamefont {A.~E.}\ \bibnamefont
  {Rastegin}}\ and\ \bibinfo {author} {\bibfnamefont {K.}~\bibnamefont
  {\ifmmode~\dot{Z}\else \.{Z}\fi{}yczkowski}},\ }\bibfield  {title} {\bibinfo
  {title} {Jarzynski equality for quantum stochastic maps},\ }\href
  {https://doi.org/10.1103/PhysRevE.89.012127} {\bibfield  {journal} {\bibinfo
  {journal} {Phys. Rev. E}\ }\textbf {\bibinfo {volume} {89}},\ \bibinfo
  {pages} {012127} (\bibinfo {year} {2014})}\BibitemShut {NoStop}%
\bibitem [{\citenamefont {Manzano}\ \emph {et~al.}(2015)\citenamefont
  {Manzano}, \citenamefont {Horowitz},\ and\ \citenamefont
  {Parrondo}}]{PhysRevE.92.032129}%
  \BibitemOpen
  \bibfield  {author} {\bibinfo {author} {\bibfnamefont {G.}~\bibnamefont
  {Manzano}}, \bibinfo {author} {\bibfnamefont {J.~M.}\ \bibnamefont
  {Horowitz}},\ and\ \bibinfo {author} {\bibfnamefont {J.~M.~R.}\ \bibnamefont
  {Parrondo}},\ }\bibfield  {title} {\bibinfo {title} {Nonequilibrium potential
  and fluctuation theorems for quantum maps},\ }\href
  {https://doi.org/10.1103/PhysRevE.92.032129} {\bibfield  {journal} {\bibinfo
  {journal} {Phys. Rev. E}\ }\textbf {\bibinfo {volume} {92}},\ \bibinfo
  {pages} {032129} (\bibinfo {year} {2015})}\BibitemShut {NoStop}%
\bibitem [{\citenamefont {Alhambra}\ \emph {et~al.}(2016)\citenamefont
  {Alhambra}, \citenamefont {Masanes}, \citenamefont {Oppenheim},\ and\
  \citenamefont {Perry}}]{PhysRevX.6.041017}%
  \BibitemOpen
  \bibfield  {author} {\bibinfo {author} {\bibfnamefont {A.~M.}\ \bibnamefont
  {Alhambra}}, \bibinfo {author} {\bibfnamefont {L.}~\bibnamefont {Masanes}},
  \bibinfo {author} {\bibfnamefont {J.}~\bibnamefont {Oppenheim}},\ and\
  \bibinfo {author} {\bibfnamefont {C.}~\bibnamefont {Perry}},\ }\bibfield
  {title} {\bibinfo {title} {Fluctuating work: From quantum thermodynamical
  identities to a second law equality},\ }\href
  {https://doi.org/10.1103/PhysRevX.6.041017} {\bibfield  {journal} {\bibinfo
  {journal} {Phys. Rev. X}\ }\textbf {\bibinfo {volume} {6}},\ \bibinfo {pages}
  {041017} (\bibinfo {year} {2016})}\BibitemShut {NoStop}%
\bibitem [{\citenamefont {Iyoda}\ \emph {et~al.}(2017)\citenamefont {Iyoda},
  \citenamefont {Kaneko},\ and\ \citenamefont {Sagawa}}]{Iyoda:2017}%
  \BibitemOpen
  \bibfield  {author} {\bibinfo {author} {\bibfnamefont {E.}~\bibnamefont
  {Iyoda}}, \bibinfo {author} {\bibfnamefont {K.}~\bibnamefont {Kaneko}},\ and\
  \bibinfo {author} {\bibfnamefont {T.}~\bibnamefont {Sagawa}},\ }\bibfield
  {title} {\bibinfo {title} {Fluctuation theorem for many-body pure quantum
  states},\ }\href {https://doi.org/10.1103/PhysRevLett.119.100601} {\bibfield
  {journal} {\bibinfo  {journal} {Phys. Rev. Lett.}\ }\textbf {\bibinfo
  {volume} {119}},\ \bibinfo {pages} {100601} (\bibinfo {year}
  {2017})}\BibitemShut {NoStop}%
\bibitem [{\citenamefont {\AA{}berg}(2018)}]{PhysRevX.8.011019}%
  \BibitemOpen
  \bibfield  {author} {\bibinfo {author} {\bibfnamefont {J.}~\bibnamefont
  {\AA{}berg}},\ }\bibfield  {title} {\bibinfo {title} {Fully quantum
  fluctuation theorems},\ }\href {https://doi.org/10.1103/PhysRevX.8.011019}
  {\bibfield  {journal} {\bibinfo  {journal} {Phys. Rev. X}\ }\textbf {\bibinfo
  {volume} {8}},\ \bibinfo {pages} {011019} (\bibinfo {year}
  {2018})}\BibitemShut {NoStop}%
\bibitem [{\citenamefont {Manzano}\ \emph {et~al.}(2018)\citenamefont
  {Manzano}, \citenamefont {Horowitz},\ and\ \citenamefont
  {Parrondo}}]{PhysRevX.8.031037}%
  \BibitemOpen
  \bibfield  {author} {\bibinfo {author} {\bibfnamefont {G.}~\bibnamefont
  {Manzano}}, \bibinfo {author} {\bibfnamefont {J.~M.}\ \bibnamefont
  {Horowitz}},\ and\ \bibinfo {author} {\bibfnamefont {J.~M.~R.}\ \bibnamefont
  {Parrondo}},\ }\bibfield  {title} {\bibinfo {title} {Quantum fluctuation
  theorems for arbitrary environments: Adiabatic and nonadiabatic entropy
  production},\ }\href {https://doi.org/10.1103/PhysRevX.8.031037} {\bibfield
  {journal} {\bibinfo  {journal} {Phys. Rev. X}\ }\textbf {\bibinfo {volume}
  {8}},\ \bibinfo {pages} {031037} (\bibinfo {year} {2018})}\BibitemShut
  {NoStop}%
\bibitem [{\citenamefont {Kawai}\ \emph {et~al.}(2007)\citenamefont {Kawai},
  \citenamefont {Parrondo},\ and\ \citenamefont {Vanden~Broeck}}]{Kawai2007}%
  \BibitemOpen
  \bibfield  {author} {\bibinfo {author} {\bibfnamefont {R.}~\bibnamefont
  {Kawai}}, \bibinfo {author} {\bibfnamefont {J.~M.~R.}\ \bibnamefont
  {Parrondo}},\ and\ \bibinfo {author} {\bibfnamefont {C.}~\bibnamefont
  {Vanden~Broeck}},\ }\bibfield  {title} {\bibinfo {title} {Dissipation: The
  phase-space perspective},\ }\href
  {https://doi.org/10.1103/PhysRevLett.98.080602} {\bibfield  {journal}
  {\bibinfo  {journal} {Phys. Rev. Lett.}\ }\textbf {\bibinfo {volume} {98}},\
  \bibinfo {pages} {080602} (\bibinfo {year} {2007})}\BibitemShut {NoStop}%
\bibitem [{\citenamefont {Parrondo}\ \emph {et~al.}(2009)\citenamefont
  {Parrondo}, \citenamefont {den Broeck},\ and\ \citenamefont
  {Kawai}}]{Parrondo2009}%
  \BibitemOpen
  \bibfield  {author} {\bibinfo {author} {\bibfnamefont {J.~M.~R.}\
  \bibnamefont {Parrondo}}, \bibinfo {author} {\bibfnamefont {C.~V.}\
  \bibnamefont {den Broeck}},\ and\ \bibinfo {author} {\bibfnamefont
  {R.}~\bibnamefont {Kawai}},\ }\bibfield  {title} {\bibinfo {title} {Entropy
  production and the arrow of time},\ }\href
  {https://doi.org/10.1088/1367-2630/11/7/073008} {\bibfield  {journal}
  {\bibinfo  {journal} {New Journal of Physics}\ }\textbf {\bibinfo {volume}
  {11}},\ \bibinfo {pages} {073008} (\bibinfo {year} {2009})}\BibitemShut
  {NoStop}%
\bibitem [{\citenamefont {Deffner}\ and\ \citenamefont {Lutz}(2011)}]{LutzPRL}%
  \BibitemOpen
  \bibfield  {author} {\bibinfo {author} {\bibfnamefont {S.}~\bibnamefont
  {Deffner}}\ and\ \bibinfo {author} {\bibfnamefont {E.}~\bibnamefont {Lutz}},\
  }\bibfield  {title} {\bibinfo {title} {Nonequilibrium entropy production for
  open quantum systems},\ }\href
  {https://doi.org/10.1103/PhysRevLett.107.140404} {\bibfield  {journal}
  {\bibinfo  {journal} {Phys. Rev. Lett.}\ }\textbf {\bibinfo {volume} {107}},\
  \bibinfo {pages} {140404} (\bibinfo {year} {2011})}\BibitemShut {NoStop}%
\bibitem [{\citenamefont {Wang}\ \emph {et~al.}(2002)\citenamefont {Wang},
  \citenamefont {Sevick}, \citenamefont {Mittag}, \citenamefont {Searles},\
  and\ \citenamefont {Evans}}]{Wang2002}%
  \BibitemOpen
  \bibfield  {author} {\bibinfo {author} {\bibfnamefont {G.~M.}\ \bibnamefont
  {Wang}}, \bibinfo {author} {\bibfnamefont {E.~M.}\ \bibnamefont {Sevick}},
  \bibinfo {author} {\bibfnamefont {E.}~\bibnamefont {Mittag}}, \bibinfo
  {author} {\bibfnamefont {D.~J.}\ \bibnamefont {Searles}},\ and\ \bibinfo
  {author} {\bibfnamefont {D.~J.}\ \bibnamefont {Evans}},\ }\bibfield  {title}
  {\bibinfo {title} {Experimental demonstration of violations of the second law
  of thermodynamics for small systems and short time scales},\ }\href
  {https://doi.org/10.1103/PhysRevLett.89.050601} {\bibfield  {journal}
  {\bibinfo  {journal} {Phys. Rev. Lett.}\ }\textbf {\bibinfo {volume} {89}},\
  \bibinfo {pages} {050601} (\bibinfo {year} {2002})}\BibitemShut {NoStop}%
\bibitem [{\citenamefont {Liphardt}\ \emph {et~al.}(2002)\citenamefont
  {Liphardt}, \citenamefont {Dumont}, \citenamefont {Smith}, \citenamefont
  {Tinoco},\ and\ \citenamefont {Bustamante}}]{Liphardt2002}%
  \BibitemOpen
  \bibfield  {author} {\bibinfo {author} {\bibfnamefont {J.}~\bibnamefont
  {Liphardt}}, \bibinfo {author} {\bibfnamefont {S.}~\bibnamefont {Dumont}},
  \bibinfo {author} {\bibfnamefont {S.~B.}\ \bibnamefont {Smith}}, \bibinfo
  {author} {\bibfnamefont {I.}~\bibnamefont {Tinoco}},\ and\ \bibinfo {author}
  {\bibfnamefont {C.}~\bibnamefont {Bustamante}},\ }\bibfield  {title}
  {\bibinfo {title} {Equilibrium information from nonequilibrium measurements
  in an experimental test of jarzynski{\textquoteright}s equality},\ }\href
  {https://doi.org/10.1126/science.1071152} {\bibfield  {journal} {\bibinfo
  {journal} {Science}\ }\textbf {\bibinfo {volume} {296}},\ \bibinfo {pages}
  {1832–1835} (\bibinfo {year} {2002})},\ \Eprint
  {https://arxiv.org/abs/https://science.sciencemag.org/content/296/5574/1832.full.pdf}
  {https://science.sciencemag.org/content/296/5574/1832.full.pdf} \BibitemShut
  {NoStop}%
\bibitem [{\citenamefont {Collin}\ \emph {et~al.}(2005)\citenamefont {Collin},
  \citenamefont {Ritort}, \citenamefont {Jarzynski}, \citenamefont {Smith},
  \citenamefont {Tinoco},\ and\ \citenamefont {Bustamante}}]{Ritort2005}%
  \BibitemOpen
  \bibfield  {author} {\bibinfo {author} {\bibfnamefont {D.}~\bibnamefont
  {Collin}}, \bibinfo {author} {\bibfnamefont {F.}~\bibnamefont {Ritort}},
  \bibinfo {author} {\bibfnamefont {C.}~\bibnamefont {Jarzynski}}, \bibinfo
  {author} {\bibfnamefont {S.~B.}\ \bibnamefont {Smith}}, \bibinfo {author}
  {\bibfnamefont {I.}~\bibnamefont {Tinoco}},\ and\ \bibinfo {author}
  {\bibfnamefont {C.}~\bibnamefont {Bustamante}},\ }\bibfield  {title}
  {\bibinfo {title} {Verification of the crooks fluctuation theorem and
  recovery of rna folding free energies},\ }\href
  {https://www.nature.com/articles/nature04061} {\bibfield  {journal} {\bibinfo
   {journal} {Nature}\ }\textbf {\bibinfo {volume} {437}},\ \bibinfo {pages}
  {231–234} (\bibinfo {year} {2005})}\BibitemShut {NoStop}%
\bibitem [{\citenamefont {Tietz}\ \emph {et~al.}(2006)\citenamefont {Tietz},
  \citenamefont {Schuler}, \citenamefont {Speck}, \citenamefont {Seifert},\
  and\ \citenamefont {Wrachtrup}}]{PhysRevLett.97.050602}%
  \BibitemOpen
  \bibfield  {author} {\bibinfo {author} {\bibfnamefont {C.}~\bibnamefont
  {Tietz}}, \bibinfo {author} {\bibfnamefont {S.}~\bibnamefont {Schuler}},
  \bibinfo {author} {\bibfnamefont {T.}~\bibnamefont {Speck}}, \bibinfo
  {author} {\bibfnamefont {U.}~\bibnamefont {Seifert}},\ and\ \bibinfo {author}
  {\bibfnamefont {J.}~\bibnamefont {Wrachtrup}},\ }\bibfield  {title} {\bibinfo
  {title} {Measurement of stochastic entropy production},\ }\href
  {https://doi.org/10.1103/PhysRevLett.97.050602} {\bibfield  {journal}
  {\bibinfo  {journal} {Phys. Rev. Lett.}\ }\textbf {\bibinfo {volume} {97}},\
  \bibinfo {pages} {050602} (\bibinfo {year} {2006})}\BibitemShut {NoStop}%
\bibitem [{\citenamefont {Toyabe}\ \emph {et~al.}(2010)\citenamefont {Toyabe},
  \citenamefont {Sagawa}, \citenamefont {Ueda}, \citenamefont {Muneyuki},\ and\
  \citenamefont {Sano}}]{Ueda2010}%
  \BibitemOpen
  \bibfield  {author} {\bibinfo {author} {\bibfnamefont {S.}~\bibnamefont
  {Toyabe}}, \bibinfo {author} {\bibfnamefont {T.}~\bibnamefont {Sagawa}},
  \bibinfo {author} {\bibfnamefont {M.}~\bibnamefont {Ueda}}, \bibinfo {author}
  {\bibfnamefont {E.}~\bibnamefont {Muneyuki}},\ and\ \bibinfo {author}
  {\bibfnamefont {M.}~\bibnamefont {Sano}},\ }\bibfield  {title} {\bibinfo
  {title} {Experimental demonstration of information-to-energy conversion and
  validation of the generalized jarzynski equality},\ }\href
  {https://www.nature.com/articles/nphys1821} {\bibfield  {journal} {\bibinfo
  {journal} {Nature Physics}\ }\textbf {\bibinfo {volume} {6}},\ \bibinfo
  {pages} {988–992} (\bibinfo {year} {2010})}\BibitemShut {NoStop}%
\bibitem [{\citenamefont {Gieseler}\ \emph {et~al.}(2014)\citenamefont
  {Gieseler}, \citenamefont {Quidant}, \citenamefont {Dellago},\ and\
  \citenamefont {Novotny}}]{NatNano.9.358}%
  \BibitemOpen
  \bibfield  {author} {\bibinfo {author} {\bibfnamefont {J.}~\bibnamefont
  {Gieseler}}, \bibinfo {author} {\bibfnamefont {R.}~\bibnamefont {Quidant}},
  \bibinfo {author} {\bibfnamefont {C.}~\bibnamefont {Dellago}},\ and\ \bibinfo
  {author} {\bibfnamefont {L.}~\bibnamefont {Novotny}},\ }\bibfield  {title}
  {\bibinfo {title} {Dynamic relaxation of a levitated nanoparticle from a
  non-equilibrium steady state},\ }\href
  {https://doi.org/10.1038/nnano.2014.40} {\bibfield  {journal} {\bibinfo
  {journal} {Nature Nanotechnology}\ }\textbf {\bibinfo {volume} {9}},\
  \bibinfo {pages} {358} (\bibinfo {year} {2014})}\BibitemShut {NoStop}%
\bibitem [{\citenamefont {Talkner}\ \emph {et~al.}(2007)\citenamefont
  {Talkner}, \citenamefont {Lutz},\ and\ \citenamefont
  {H\"anggi}}]{PhysRevE.75.050102}%
  \BibitemOpen
  \bibfield  {author} {\bibinfo {author} {\bibfnamefont {P.}~\bibnamefont
  {Talkner}}, \bibinfo {author} {\bibfnamefont {E.}~\bibnamefont {Lutz}},\ and\
  \bibinfo {author} {\bibfnamefont {P.}~\bibnamefont {H\"anggi}},\ }\bibfield
  {title} {\bibinfo {title} {Fluctuation theorems: Work is not an observable},\
  }\href {https://doi.org/10.1103/PhysRevE.75.050102} {\bibfield  {journal}
  {\bibinfo  {journal} {Phys. Rev. E}\ }\textbf {\bibinfo {volume} {75}},\
  \bibinfo {pages} {050102(R)} (\bibinfo {year} {2007})}\BibitemShut {NoStop}%
\bibitem [{\citenamefont {Watanabe}\ \emph {et~al.}(2014)\citenamefont
  {Watanabe}, \citenamefont {Venkatesh},\ and\ \citenamefont
  {Talkner}}]{Watanabe_2014}%
  \BibitemOpen
  \bibfield  {author} {\bibinfo {author} {\bibfnamefont {G.}~\bibnamefont
  {Watanabe}}, \bibinfo {author} {\bibfnamefont {B.~P.}\ \bibnamefont
  {Venkatesh}},\ and\ \bibinfo {author} {\bibfnamefont {P.}~\bibnamefont
  {Talkner}},\ }\bibfield  {title} {\bibinfo {title} {Generalized energy
  measurements and modified transient quantum fluctuation theorems},\ }\href
  {https://doi.org/10.1103/PhysRevE.89.052116} {\bibfield  {journal} {\bibinfo
  {journal} {Phys. Rev. E}\ }\textbf {\bibinfo {volume} {89}},\ \bibinfo
  {pages} {052116} (\bibinfo {year} {2014})}\BibitemShut {NoStop}%
\bibitem [{\citenamefont {Ito}\ \emph {et~al.}(2019)\citenamefont {Ito},
  \citenamefont {Talkner}, \citenamefont {Venkatesh},\ and\ \citenamefont
  {Watanabe}}]{Talkner_2019}%
  \BibitemOpen
  \bibfield  {author} {\bibinfo {author} {\bibfnamefont {K.}~\bibnamefont
  {Ito}}, \bibinfo {author} {\bibfnamefont {P.}~\bibnamefont {Talkner}},
  \bibinfo {author} {\bibfnamefont {B.~P.}\ \bibnamefont {Venkatesh}},\ and\
  \bibinfo {author} {\bibfnamefont {G.}~\bibnamefont {Watanabe}},\ }\bibfield
  {title} {\bibinfo {title} {Generalized energy measurements and quantum work
  compatible with fluctuation theorems},\ }\href
  {https://doi.org/10.1103/PhysRevA.99.032117} {\bibfield  {journal} {\bibinfo
  {journal} {Phys. Rev. A}\ }\textbf {\bibinfo {volume} {99}},\ \bibinfo
  {pages} {032117} (\bibinfo {year} {2019})}\BibitemShut {NoStop}%
\bibitem [{\citenamefont {Debarba}\ \emph {et~al.}(2019)\citenamefont
  {Debarba}, \citenamefont {Manzano}, \citenamefont {Guryanova}, \citenamefont
  {Huber},\ and\ \citenamefont {Friis}}]{Debarba_2019}%
  \BibitemOpen
  \bibfield  {author} {\bibinfo {author} {\bibfnamefont {T.}~\bibnamefont
  {Debarba}}, \bibinfo {author} {\bibfnamefont {G.}~\bibnamefont {Manzano}},
  \bibinfo {author} {\bibfnamefont {Y.}~\bibnamefont {Guryanova}}, \bibinfo
  {author} {\bibfnamefont {M.}~\bibnamefont {Huber}},\ and\ \bibinfo {author}
  {\bibfnamefont {N.}~\bibnamefont {Friis}},\ }\bibfield  {title} {\bibinfo
  {title} {Work estimation and work fluctuations in the presence of non-ideal
  measurements},\ }\href {https://doi.org/10.1088/1367-2630/ab4d9d} {\bibfield
  {journal} {\bibinfo  {journal} {New Journal of Physics}\ }\textbf {\bibinfo
  {volume} {21}},\ \bibinfo {pages} {113002} (\bibinfo {year}
  {2019})}\BibitemShut {NoStop}%
\bibitem [{\citenamefont {Solinas}\ and\ \citenamefont
  {Gasparinetti}(2016)}]{Solinas:2016}%
  \BibitemOpen
  \bibfield  {author} {\bibinfo {author} {\bibfnamefont {P.}~\bibnamefont
  {Solinas}}\ and\ \bibinfo {author} {\bibfnamefont {S.}~\bibnamefont
  {Gasparinetti}},\ }\bibfield  {title} {\bibinfo {title} {Probing quantum
  interference effects in the work distribution},\ }\href
  {https://doi.org/10.1103/PhysRevA.94.052103} {\bibfield  {journal} {\bibinfo
  {journal} {Phys. Rev. A}\ }\textbf {\bibinfo {volume} {94}},\ \bibinfo
  {pages} {052103} (\bibinfo {year} {2016})}\BibitemShut {NoStop}%
\bibitem [{\citenamefont {Perarnau-Llobet}\ \emph {et~al.}(2017)\citenamefont
  {Perarnau-Llobet}, \citenamefont {B\"aumer}, \citenamefont {Hovhannisyan},
  \citenamefont {Huber},\ and\ \citenamefont {Acin}}]{PhysRevLett.118.070601}%
  \BibitemOpen
  \bibfield  {author} {\bibinfo {author} {\bibfnamefont {M.}~\bibnamefont
  {Perarnau-Llobet}}, \bibinfo {author} {\bibfnamefont {E.}~\bibnamefont
  {B\"aumer}}, \bibinfo {author} {\bibfnamefont {K.~V.}\ \bibnamefont
  {Hovhannisyan}}, \bibinfo {author} {\bibfnamefont {M.}~\bibnamefont
  {Huber}},\ and\ \bibinfo {author} {\bibfnamefont {A.}~\bibnamefont {Acin}},\
  }\bibfield  {title} {\bibinfo {title} {No-go theorem for the characterization
  of work fluctuations in coherent quantum systems},\ }\href
  {https://doi.org/10.1103/PhysRevLett.118.070601} {\bibfield  {journal}
  {\bibinfo  {journal} {Phys. Rev. Lett.}\ }\textbf {\bibinfo {volume} {118}},\
  \bibinfo {pages} {070601} (\bibinfo {year} {2017})}\BibitemShut {NoStop}%
\bibitem [{\citenamefont {Lostaglio}(2018)}]{PhysRevLett.120.040602}%
  \BibitemOpen
  \bibfield  {author} {\bibinfo {author} {\bibfnamefont {M.}~\bibnamefont
  {Lostaglio}},\ }\bibfield  {title} {\bibinfo {title} {Quantum fluctuation
  theorems, contextuality, and work quasiprobabilities},\ }\href
  {https://doi.org/10.1103/PhysRevLett.120.040602} {\bibfield  {journal}
  {\bibinfo  {journal} {Phys. Rev. Lett.}\ }\textbf {\bibinfo {volume} {120}},\
  \bibinfo {pages} {040602} (\bibinfo {year} {2018})}\BibitemShut {NoStop}%
\bibitem [{\citenamefont {Sone}\ \emph {et~al.}(2020)\citenamefont {Sone},
  \citenamefont {Liu},\ and\ \citenamefont {Cappellaro}}]{Sone_2020}%
  \BibitemOpen
  \bibfield  {author} {\bibinfo {author} {\bibfnamefont {A.}~\bibnamefont
  {Sone}}, \bibinfo {author} {\bibfnamefont {Y.-X.}\ \bibnamefont {Liu}},\ and\
  \bibinfo {author} {\bibfnamefont {P.}~\bibnamefont {Cappellaro}},\ }\bibfield
   {title} {\bibinfo {title} {Quantum jarzynski equality in open quantum
  systems from the one-time measurement scheme},\ }\href
  {https://doi.org/10.1103/PhysRevLett.125.060602} {\bibfield  {journal}
  {\bibinfo  {journal} {Phys. Rev. Lett.}\ }\textbf {\bibinfo {volume} {125}},\
  \bibinfo {pages} {060602} (\bibinfo {year} {2020})}\BibitemShut {NoStop}%
\bibitem [{\citenamefont {Beyer}\ \emph {et~al.}(2020)\citenamefont {Beyer},
  \citenamefont {Luoma},\ and\ \citenamefont {Strunz}}]{Beyer_2020}%
  \BibitemOpen
  \bibfield  {author} {\bibinfo {author} {\bibfnamefont {K.}~\bibnamefont
  {Beyer}}, \bibinfo {author} {\bibfnamefont {K.}~\bibnamefont {Luoma}},\ and\
  \bibinfo {author} {\bibfnamefont {W.~T.}\ \bibnamefont {Strunz}},\ }\bibfield
   {title} {\bibinfo {title} {Work as an external quantum observable and an
  operational quantum work fluctuation theorem},\ }\href
  {https://doi.org/10.1103/PhysRevResearch.2.033508} {\bibfield  {journal}
  {\bibinfo  {journal} {Phys. Rev. Research}\ }\textbf {\bibinfo {volume}
  {2}},\ \bibinfo {pages} {033508} (\bibinfo {year} {2020})}\BibitemShut
  {NoStop}%
\bibitem [{\citenamefont {Micadei}\ \emph {et~al.}(2020)\citenamefont
  {Micadei}, \citenamefont {Landi},\ and\ \citenamefont {Lutz}}]{Micadei_2020}%
  \BibitemOpen
  \bibfield  {author} {\bibinfo {author} {\bibfnamefont {K.}~\bibnamefont
  {Micadei}}, \bibinfo {author} {\bibfnamefont {G.~T.}\ \bibnamefont {Landi}},\
  and\ \bibinfo {author} {\bibfnamefont {E.}~\bibnamefont {Lutz}},\ }\bibfield
  {title} {\bibinfo {title} {Quantum fluctuation theorems beyond two-point
  measurements},\ }\href {https://doi.org/10.1103/PhysRevLett.124.090602}
  {\bibfield  {journal} {\bibinfo  {journal} {Phys. Rev. Lett.}\ }\textbf
  {\bibinfo {volume} {124}},\ \bibinfo {pages} {090602} (\bibinfo {year}
  {2020})}\BibitemShut {NoStop}%
\bibitem [{\citenamefont {An}\ \emph {et~al.}(2015)\citenamefont {An},
  \citenamefont {Zhang}, \citenamefont {Um}, \citenamefont {Lv}, \citenamefont
  {Lu}, \citenamefont {Zhang}, \citenamefont {Yin}, \citenamefont {Quan},\ and\
  \citenamefont {Kim}}]{Kim2015}%
  \BibitemOpen
  \bibfield  {author} {\bibinfo {author} {\bibfnamefont {S.}~\bibnamefont
  {An}}, \bibinfo {author} {\bibfnamefont {J.-N.}\ \bibnamefont {Zhang}},
  \bibinfo {author} {\bibfnamefont {M.}~\bibnamefont {Um}}, \bibinfo {author}
  {\bibfnamefont {D.}~\bibnamefont {Lv}}, \bibinfo {author} {\bibfnamefont
  {Y.}~\bibnamefont {Lu}}, \bibinfo {author} {\bibfnamefont {J.}~\bibnamefont
  {Zhang}}, \bibinfo {author} {\bibfnamefont {Z.-Q.}\ \bibnamefont {Yin}},
  \bibinfo {author} {\bibfnamefont {H.~T.}\ \bibnamefont {Quan}},\ and\
  \bibinfo {author} {\bibfnamefont {K.}~\bibnamefont {Kim}},\ }\bibfield
  {title} {\bibinfo {title} {Experimental test of the quantum jarzynski
  equality with a trapped-ion system},\ }\href
  {https://www.nature.com/articles/nphys3197} {\bibfield  {journal} {\bibinfo
  {journal} {Nature Physics}\ }\textbf {\bibinfo {volume} {11}},\ \bibinfo
  {pages} {193–199} (\bibinfo {year} {2015})}\BibitemShut {NoStop}%
\bibitem [{\citenamefont {Xiong}\ \emph {et~al.}(2018)\citenamefont {Xiong},
  \citenamefont {Yan}, \citenamefont {Zhou}, \citenamefont {Rehan},
  \citenamefont {Liang}, \citenamefont {Chen}, \citenamefont {Yang},
  \citenamefont {Ma}, \citenamefont {Feng},\ and\ \citenamefont
  {Vedral}}]{PhysRevLett.120.010601}%
  \BibitemOpen
  \bibfield  {author} {\bibinfo {author} {\bibfnamefont {T.~P.}\ \bibnamefont
  {Xiong}}, \bibinfo {author} {\bibfnamefont {L.~L.}\ \bibnamefont {Yan}},
  \bibinfo {author} {\bibfnamefont {F.}~\bibnamefont {Zhou}}, \bibinfo {author}
  {\bibfnamefont {K.}~\bibnamefont {Rehan}}, \bibinfo {author} {\bibfnamefont
  {D.~F.}\ \bibnamefont {Liang}}, \bibinfo {author} {\bibfnamefont
  {L.}~\bibnamefont {Chen}}, \bibinfo {author} {\bibfnamefont {W.~L.}\
  \bibnamefont {Yang}}, \bibinfo {author} {\bibfnamefont {Z.~H.}\ \bibnamefont
  {Ma}}, \bibinfo {author} {\bibfnamefont {M.}~\bibnamefont {Feng}},\ and\
  \bibinfo {author} {\bibfnamefont {V.}~\bibnamefont {Vedral}},\ }\bibfield
  {title} {\bibinfo {title} {Experimental verification of a jarzynski-related
  information-theoretic equality by a single trapped ion},\ }\href
  {https://doi.org/10.1103/PhysRevLett.120.010601} {\bibfield  {journal}
  {\bibinfo  {journal} {Phys. Rev. Lett.}\ }\textbf {\bibinfo {volume} {120}},\
  \bibinfo {pages} {010601} (\bibinfo {year} {2018})}\BibitemShut {NoStop}%
\bibitem [{\citenamefont {Campisi}\ and\ \citenamefont
  {H\"anggi}(2018)}]{PhysRevLett.121.088901}%
  \BibitemOpen
  \bibfield  {author} {\bibinfo {author} {\bibfnamefont {M.}~\bibnamefont
  {Campisi}}\ and\ \bibinfo {author} {\bibfnamefont {P.}~\bibnamefont
  {H\"anggi}},\ }\bibfield  {title} {\bibinfo {title} {Comment on
  ``experimental verification of a jarzynski-related information-theoretic
  equality by a single trapped ion''},\ }\href
  {https://doi.org/10.1103/PhysRevLett.121.088901} {\bibfield  {journal}
  {\bibinfo  {journal} {Phys. Rev. Lett.}\ }\textbf {\bibinfo {volume} {121}},\
  \bibinfo {pages} {088901} (\bibinfo {year} {2018})}\BibitemShut {NoStop}%
\bibitem [{\citenamefont {Zhang}\ \emph {et~al.}(2018)\citenamefont {Zhang},
  \citenamefont {Wang}, \citenamefont {Xiang}, \citenamefont {Jia},
  \citenamefont {Duan}, \citenamefont {Cai}, \citenamefont {Zhan},
  \citenamefont {Zong}, \citenamefont {Wu}, \citenamefont {Sun}, \citenamefont
  {Yin},\ and\ \citenamefont {Guo}}]{Zhang_2018}%
  \BibitemOpen
  \bibfield  {author} {\bibinfo {author} {\bibfnamefont {Z.}~\bibnamefont
  {Zhang}}, \bibinfo {author} {\bibfnamefont {T.}~\bibnamefont {Wang}},
  \bibinfo {author} {\bibfnamefont {L.}~\bibnamefont {Xiang}}, \bibinfo
  {author} {\bibfnamefont {Z.}~\bibnamefont {Jia}}, \bibinfo {author}
  {\bibfnamefont {P.}~\bibnamefont {Duan}}, \bibinfo {author} {\bibfnamefont
  {W.}~\bibnamefont {Cai}}, \bibinfo {author} {\bibfnamefont {Z.}~\bibnamefont
  {Zhan}}, \bibinfo {author} {\bibfnamefont {Z.}~\bibnamefont {Zong}}, \bibinfo
  {author} {\bibfnamefont {J.}~\bibnamefont {Wu}}, \bibinfo {author}
  {\bibfnamefont {L.}~\bibnamefont {Sun}}, \bibinfo {author} {\bibfnamefont
  {Y.}~\bibnamefont {Yin}},\ and\ \bibinfo {author} {\bibfnamefont
  {G.}~\bibnamefont {Guo}},\ }\bibfield  {title} {\bibinfo {title}
  {Experimental demonstration of work fluctuations along a shortcut to
  adiabaticity with a superconducting xmon qubit},\ }\href
  {https://doi.org/10.1088/1367-2630/aad4e7} {\bibfield  {journal} {\bibinfo
  {journal} {New Journal of Physics}\ }\textbf {\bibinfo {volume} {20}},\
  \bibinfo {pages} {085001} (\bibinfo {year} {2018})}\BibitemShut {NoStop}%
\bibitem [{\citenamefont {Wu}\ \emph {et~al.}(2019)\citenamefont {Wu},
  \citenamefont {Yuan}, \citenamefont {Xiang}, \citenamefont {Li},
  \citenamefont {Guo},\ and\ \citenamefont {Perarnau-Llobet}}]{Wueaav4944}%
  \BibitemOpen
  \bibfield  {author} {\bibinfo {author} {\bibfnamefont {K.-D.}\ \bibnamefont
  {Wu}}, \bibinfo {author} {\bibfnamefont {Y.}~\bibnamefont {Yuan}}, \bibinfo
  {author} {\bibfnamefont {G.-Y.}\ \bibnamefont {Xiang}}, \bibinfo {author}
  {\bibfnamefont {C.-F.}\ \bibnamefont {Li}}, \bibinfo {author} {\bibfnamefont
  {G.-C.}\ \bibnamefont {Guo}},\ and\ \bibinfo {author} {\bibfnamefont
  {M.}~\bibnamefont {Perarnau-Llobet}},\ }\bibfield  {title} {\bibinfo {title}
  {Experimentally reducing the quantum measurement back action in work
  distributions by a collective measurement},\ }\bibfield  {journal} {\bibinfo
  {journal} {Science Advances}\ }\textbf {\bibinfo {volume} {5}},\ \href
  {https://doi.org/10.1126/sciadv.aav4944} {10.1126/sciadv.aav4944} (\bibinfo
  {year} {2019}),\ \Eprint
  {https://arxiv.org/abs/https://advances.sciencemag.org/content/5/3/eaav4944.full.pdf}
  {https://advances.sciencemag.org/content/5/3/eaav4944.full.pdf} \BibitemShut
  {NoStop}%
\bibitem [{\citenamefont {Dorner}\ \emph {et~al.}(2013)\citenamefont {Dorner},
  \citenamefont {Clark}, \citenamefont {Heaney}, \citenamefont {Fazio},
  \citenamefont {Goold},\ and\ \citenamefont
  {Vedral}}]{PhysRevLett.110.230601}%
  \BibitemOpen
  \bibfield  {author} {\bibinfo {author} {\bibfnamefont {R.}~\bibnamefont
  {Dorner}}, \bibinfo {author} {\bibfnamefont {S.~R.}\ \bibnamefont {Clark}},
  \bibinfo {author} {\bibfnamefont {L.}~\bibnamefont {Heaney}}, \bibinfo
  {author} {\bibfnamefont {R.}~\bibnamefont {Fazio}}, \bibinfo {author}
  {\bibfnamefont {J.}~\bibnamefont {Goold}},\ and\ \bibinfo {author}
  {\bibfnamefont {V.}~\bibnamefont {Vedral}},\ }\bibfield  {title} {\bibinfo
  {title} {Extracting quantum work statistics and fluctuation theorems by
  single-qubit interferometry},\ }\href
  {https://doi.org/10.1103/PhysRevLett.110.230601} {\bibfield  {journal}
  {\bibinfo  {journal} {Phys. Rev. Lett.}\ }\textbf {\bibinfo {volume} {110}},\
  \bibinfo {pages} {230601} (\bibinfo {year} {2013})}\BibitemShut {NoStop}%
\bibitem [{\citenamefont {Mazzola}\ \emph {et~al.}(2013)\citenamefont
  {Mazzola}, \citenamefont {De~Chiara},\ and\ \citenamefont
  {Paternostro}}]{PhysRevLett.110.230602}%
  \BibitemOpen
  \bibfield  {author} {\bibinfo {author} {\bibfnamefont {L.}~\bibnamefont
  {Mazzola}}, \bibinfo {author} {\bibfnamefont {G.}~\bibnamefont {De~Chiara}},\
  and\ \bibinfo {author} {\bibfnamefont {M.}~\bibnamefont {Paternostro}},\
  }\bibfield  {title} {\bibinfo {title} {Measuring the characteristic function
  of the work distribution},\ }\href
  {https://doi.org/10.1103/PhysRevLett.110.230602} {\bibfield  {journal}
  {\bibinfo  {journal} {Phys. Rev. Lett.}\ }\textbf {\bibinfo {volume} {110}},\
  \bibinfo {pages} {230602} (\bibinfo {year} {2013})}\BibitemShut {NoStop}%
\bibitem [{\citenamefont {Batalh\~ao}\ \emph {et~al.}(2014)\citenamefont
  {Batalh\~ao}, \citenamefont {Souza}, \citenamefont {Mazzola}, \citenamefont
  {Auccaise}, \citenamefont {Sarthour}, \citenamefont {Oliveira}, \citenamefont
  {Goold}, \citenamefont {De~Chiara}, \citenamefont {Paternostro},\ and\
  \citenamefont {Serra}}]{Serra2014}%
  \BibitemOpen
  \bibfield  {author} {\bibinfo {author} {\bibfnamefont {T.~B.}\ \bibnamefont
  {Batalh\~ao}}, \bibinfo {author} {\bibfnamefont {A.~M.}\ \bibnamefont
  {Souza}}, \bibinfo {author} {\bibfnamefont {L.}~\bibnamefont {Mazzola}},
  \bibinfo {author} {\bibfnamefont {R.}~\bibnamefont {Auccaise}}, \bibinfo
  {author} {\bibfnamefont {R.~S.}\ \bibnamefont {Sarthour}}, \bibinfo {author}
  {\bibfnamefont {I.~S.}\ \bibnamefont {Oliveira}}, \bibinfo {author}
  {\bibfnamefont {J.}~\bibnamefont {Goold}}, \bibinfo {author} {\bibfnamefont
  {G.}~\bibnamefont {De~Chiara}}, \bibinfo {author} {\bibfnamefont
  {M.}~\bibnamefont {Paternostro}},\ and\ \bibinfo {author} {\bibfnamefont
  {R.~M.}\ \bibnamefont {Serra}},\ }\bibfield  {title} {\bibinfo {title}
  {Experimental reconstruction of work distribution and study of fluctuation
  relations in a closed quantum system},\ }\href
  {https://doi.org/10.1103/PhysRevLett.113.140601} {\bibfield  {journal}
  {\bibinfo  {journal} {Phys. Rev. Lett.}\ }\textbf {\bibinfo {volume} {113}},\
  \bibinfo {pages} {140601} (\bibinfo {year} {2014})}\BibitemShut {NoStop}%
\bibitem [{\citenamefont {Batalh\~ao}\ \emph {et~al.}(2015)\citenamefont
  {Batalh\~ao}, \citenamefont {Souza}, \citenamefont {Sarthour}, \citenamefont
  {Oliveira}, \citenamefont {Paternostro}, \citenamefont {Lutz},\ and\
  \citenamefont {Serra}}]{Serra2015}%
  \BibitemOpen
  \bibfield  {author} {\bibinfo {author} {\bibfnamefont {T.~B.}\ \bibnamefont
  {Batalh\~ao}}, \bibinfo {author} {\bibfnamefont {A.~M.}\ \bibnamefont
  {Souza}}, \bibinfo {author} {\bibfnamefont {R.~S.}\ \bibnamefont {Sarthour}},
  \bibinfo {author} {\bibfnamefont {I.~S.}\ \bibnamefont {Oliveira}}, \bibinfo
  {author} {\bibfnamefont {M.}~\bibnamefont {Paternostro}}, \bibinfo {author}
  {\bibfnamefont {E.}~\bibnamefont {Lutz}},\ and\ \bibinfo {author}
  {\bibfnamefont {R.~M.}\ \bibnamefont {Serra}},\ }\bibfield  {title} {\bibinfo
  {title} {Irreversibility and the arrow of time in a quenched quantum
  system},\ }\href {https://doi.org/10.1103/PhysRevLett.115.190601} {\bibfield
  {journal} {\bibinfo  {journal} {Phys. Rev. Lett.}\ }\textbf {\bibinfo
  {volume} {115}},\ \bibinfo {pages} {190601} (\bibinfo {year}
  {2015})}\BibitemShut {NoStop}%
\bibitem [{\citenamefont {Roncaglia}\ \emph {et~al.}(2014)\citenamefont
  {Roncaglia}, \citenamefont {Cerisola},\ and\ \citenamefont {Paz}}]{Paz2014}%
  \BibitemOpen
  \bibfield  {author} {\bibinfo {author} {\bibfnamefont {A.~J.}\ \bibnamefont
  {Roncaglia}}, \bibinfo {author} {\bibfnamefont {F.}~\bibnamefont
  {Cerisola}},\ and\ \bibinfo {author} {\bibfnamefont {J.~P.}\ \bibnamefont
  {Paz}},\ }\bibfield  {title} {\bibinfo {title} {Work measurement as a
  generalized quantum measurement},\ }\href
  {https://doi.org/10.1103/PhysRevLett.113.250601} {\bibfield  {journal}
  {\bibinfo  {journal} {Phys. Rev. Lett.}\ }\textbf {\bibinfo {volume} {113}},\
  \bibinfo {pages} {250601} (\bibinfo {year} {2014})}\BibitemShut {NoStop}%
\bibitem [{\citenamefont {Chiara}\ \emph {et~al.}(2015)\citenamefont {Chiara},
  \citenamefont {Roncaglia},\ and\ \citenamefont {Paz}}]{Chiara_2015}%
  \BibitemOpen
  \bibfield  {author} {\bibinfo {author} {\bibfnamefont {G.~D.}\ \bibnamefont
  {Chiara}}, \bibinfo {author} {\bibfnamefont {A.~J.}\ \bibnamefont
  {Roncaglia}},\ and\ \bibinfo {author} {\bibfnamefont {J.~P.}\ \bibnamefont
  {Paz}},\ }\bibfield  {title} {\bibinfo {title} {Measuring work and heat in
  ultracold quantum gases},\ }\href
  {https://doi.org/10.1088/1367-2630/17/3/035004} {\bibfield  {journal}
  {\bibinfo  {journal} {New Journal of Physics}\ }\textbf {\bibinfo {volume}
  {17}},\ \bibinfo {pages} {035004} (\bibinfo {year} {2015})}\BibitemShut
  {NoStop}%
\bibitem [{\citenamefont {Talkner}\ and\ \citenamefont
  {H\"anggi}(2016)}]{Talkner_2016}%
  \BibitemOpen
  \bibfield  {author} {\bibinfo {author} {\bibfnamefont {P.}~\bibnamefont
  {Talkner}}\ and\ \bibinfo {author} {\bibfnamefont {P.}~\bibnamefont
  {H\"anggi}},\ }\bibfield  {title} {\bibinfo {title} {Aspects of quantum
  work},\ }\href {https://doi.org/10.1103/PhysRevE.93.022131} {\bibfield
  {journal} {\bibinfo  {journal} {Phys. Rev. E}\ }\textbf {\bibinfo {volume}
  {93}},\ \bibinfo {pages} {022131} (\bibinfo {year} {2016})}\BibitemShut
  {NoStop}%
\bibitem [{\citenamefont {Cerisola}\ \emph {et~al.}(2017)\citenamefont
  {Cerisola}, \citenamefont {Margalit}, \citenamefont {Machluf}, \citenamefont
  {Roncaglia}, \citenamefont {Paz},\ and\ \citenamefont
  {Folman}}]{Cersiola2017}%
  \BibitemOpen
  \bibfield  {author} {\bibinfo {author} {\bibfnamefont {F.}~\bibnamefont
  {Cerisola}}, \bibinfo {author} {\bibfnamefont {Y.}~\bibnamefont {Margalit}},
  \bibinfo {author} {\bibfnamefont {S.}~\bibnamefont {Machluf}}, \bibinfo
  {author} {\bibfnamefont {A.~J.}\ \bibnamefont {Roncaglia}}, \bibinfo {author}
  {\bibfnamefont {J.~P.}\ \bibnamefont {Paz}},\ and\ \bibinfo {author}
  {\bibfnamefont {R.}~\bibnamefont {Folman}},\ }\bibfield  {title} {\bibinfo
  {title} {Using a quantum work meter to test non-equilibrium fluctuation
  theorems},\ }\href {https://www.nature.com/articles/s41467-017-01308-7}
  {\bibfield  {journal} {\bibinfo  {journal} {Nature Communications}\ }\textbf
  {\bibinfo {volume} {8}},\ \bibinfo {pages} {1241} (\bibinfo {year}
  {2017})}\BibitemShut {NoStop}%
\bibitem [{\citenamefont {{Rubino}}\ \emph {et~al.}(2021)\citenamefont
  {{Rubino}}, \citenamefont {{Manzano}},\ and\ \citenamefont
  {{Brukner}}}]{Rubino:2020}%
  \BibitemOpen
  \bibfield  {author} {\bibinfo {author} {\bibfnamefont {G.}~\bibnamefont
  {{Rubino}}}, \bibinfo {author} {\bibfnamefont {G.}~\bibnamefont
  {{Manzano}}},\ and\ \bibinfo {author} {\bibfnamefont {{\v{C}}.}~\bibnamefont
  {{Brukner}}},\ }\bibfield  {title} {\bibinfo {title} {Quantum superposition
  of thermodynamic evolutions with opposing time’s arrows},\ }\href
  {https://doi.org/10.1038/s42005-021-00759-1} {\bibfield  {journal} {\bibinfo
  {journal} {Comm. Phys.}\ }\textbf {\bibinfo {volume} {4}},\ \bibinfo {pages}
  {251} (\bibinfo {year} {2021})}\BibitemShut {NoStop}%
\bibitem [{Note1()}]{Note1}%
  \BibitemOpen
  \bibinfo {note} {The same result can be extended to Hamiltonians which are
  not invariant under time-reversal. In such a case, the initial state of the
  time-reversal process needs to incorporate the broken symmetry, that is,
  $\protect \tilde {\rho }_0^\protect \mathrm {th}= \Theta \protect \, \rho
  _\tau ^{\protect \mathrm {th}}$}\BibitemShut {NoStop}%
\bibitem [{\citenamefont {Englert}(1996)}]{PhysRevLett.77.2154}%
  \BibitemOpen
  \bibfield  {author} {\bibinfo {author} {\bibfnamefont {B.-G.}\ \bibnamefont
  {Englert}},\ }\bibfield  {title} {\bibinfo {title} {Fringe visibility and
  which-way information: An inequality},\ }\href
  {https://doi.org/10.1103/PhysRevLett.77.2154} {\bibfield  {journal} {\bibinfo
   {journal} {Phys. Rev. Lett.}\ }\textbf {\bibinfo {volume} {77}},\ \bibinfo
  {pages} {2154} (\bibinfo {year} {1996})}\BibitemShut {NoStop}%
\bibitem [{\citenamefont {Audenaert}\ and\ \citenamefont
  {Eisert}(2005)}]{doi:10.1063/1.2044667}%
  \BibitemOpen
  \bibfield  {author} {\bibinfo {author} {\bibfnamefont {K.~M.~R.}\
  \bibnamefont {Audenaert}}\ and\ \bibinfo {author} {\bibfnamefont
  {J.}~\bibnamefont {Eisert}},\ }\bibfield  {title} {\bibinfo {title}
  {Continuity bounds on the quantum relative entropy},\ }\href
  {https://doi.org/10.1063/1.2044667} {\bibfield  {journal} {\bibinfo
  {journal} {Journal of Mathematical Physics}\ }\textbf {\bibinfo {volume}
  {46}},\ \bibinfo {pages} {102104} (\bibinfo {year} {2005})},\ \Eprint
  {https://arxiv.org/abs/https://doi.org/10.1063/1.2044667}
  {https://doi.org/10.1063/1.2044667} \BibitemShut {NoStop}%
\bibitem [{\citenamefont {Friis}\ \emph {et~al.}(2014)\citenamefont {Friis},
  \citenamefont {Dunjko}, \citenamefont {D{\"u}r},\ and\ \citenamefont
  {Briegel}}]{friis2014implementing}%
  \BibitemOpen
  \bibfield  {author} {\bibinfo {author} {\bibfnamefont {N.}~\bibnamefont
  {Friis}}, \bibinfo {author} {\bibfnamefont {V.}~\bibnamefont {Dunjko}},
  \bibinfo {author} {\bibfnamefont {W.}~\bibnamefont {D{\"u}r}},\ and\ \bibinfo
  {author} {\bibfnamefont {H.~J.}\ \bibnamefont {Briegel}},\ }\bibfield
  {title} {\bibinfo {title} {Implementing quantum control for unknown
  subroutines},\ }\href@noop {} {\bibfield  {journal} {\bibinfo  {journal}
  {Physical Review A}\ }\textbf {\bibinfo {volume} {89}},\ \bibinfo {pages}
  {030303(R)} (\bibinfo {year} {2014})}\BibitemShut {NoStop}%
\bibitem [{\citenamefont {Gooch}\ and\ \citenamefont
  {Tarry}(1975)}]{gooch1975optical}%
  \BibitemOpen
  \bibfield  {author} {\bibinfo {author} {\bibfnamefont {C.}~\bibnamefont
  {Gooch}}\ and\ \bibinfo {author} {\bibfnamefont {H.}~\bibnamefont {Tarry}},\
  }\bibfield  {title} {\bibinfo {title} {The optical properties of twisted
  nematic liquid crystal structures with twist angles $\leq$ 90 degrees},\
  }\href@noop {} {\bibfield  {journal} {\bibinfo  {journal} {Journal of Physics
  D: Applied Physics}\ }\textbf {\bibinfo {volume} {8}},\ \bibinfo {pages}
  {1575} (\bibinfo {year} {1975})}\BibitemShut {NoStop}%
\bibitem [{\citenamefont {Takatoh}\ \emph {et~al.}(2012)\citenamefont
  {Takatoh}, \citenamefont {Harima}, \citenamefont {Kaname}, \citenamefont
  {Shinohara},\ and\ \citenamefont {Akimoto}}]{takatoh2012fast}%
  \BibitemOpen
  \bibfield  {author} {\bibinfo {author} {\bibfnamefont {K.}~\bibnamefont
  {Takatoh}}, \bibinfo {author} {\bibfnamefont {A.}~\bibnamefont {Harima}},
  \bibinfo {author} {\bibfnamefont {Y.}~\bibnamefont {Kaname}}, \bibinfo
  {author} {\bibfnamefont {K.}~\bibnamefont {Shinohara}},\ and\ \bibinfo
  {author} {\bibfnamefont {M.}~\bibnamefont {Akimoto}},\ }\bibfield  {title}
  {\bibinfo {title} {Fast-response twisted nematic liquid crystal displays with
  ultrashort pitch liquid crystalline materials},\ }\href@noop {} {\bibfield
  {journal} {\bibinfo  {journal} {Liquid Crystals}\ }\textbf {\bibinfo {volume}
  {39}},\ \bibinfo {pages} {715} (\bibinfo {year} {2012})}\BibitemShut
  {NoStop}%
\bibitem [{\citenamefont {Park}\ \emph {et~al.}(2020)\citenamefont {Park},
  \citenamefont {Nha}, \citenamefont {Kim},\ and\ \citenamefont
  {Vedral}}]{PhysRevE.101.052128}%
  \BibitemOpen
  \bibfield  {author} {\bibinfo {author} {\bibfnamefont {J.~J.}\ \bibnamefont
  {Park}}, \bibinfo {author} {\bibfnamefont {H.}~\bibnamefont {Nha}}, \bibinfo
  {author} {\bibfnamefont {S.~W.}\ \bibnamefont {Kim}},\ and\ \bibinfo {author}
  {\bibfnamefont {V.}~\bibnamefont {Vedral}},\ }\bibfield  {title} {\bibinfo
  {title} {Information fluctuation theorem for an open quantum bipartite
  system},\ }\href {https://doi.org/10.1103/PhysRevE.101.052128} {\bibfield
  {journal} {\bibinfo  {journal} {Phys. Rev. E}\ }\textbf {\bibinfo {volume}
  {101}},\ \bibinfo {pages} {052128} (\bibinfo {year} {2020})}\BibitemShut
  {NoStop}%
\end{thebibliography}%

\end{document}